\documentclass[manuscript]{aastex}
\usepackage{psfig}

\begin{document}

\title{Tip of the Red Giant Branch Distances. I. Optimization of a Maximum
Likelihood Algorithm}

\author{Dmitry Makarov\altaffilmark{1,2,3},\ Lidia Makarova\altaffilmark{1,2},\ Luca Rizzi and R. Brent Tully}
\affil{Institute for Astronomy, University of Hawaii, 2680 Woodlawn Drive,
 Honolulu, HI 96822}

\and

\author{Andrew E. Dolphin}
\affil{Steward Observatory, University of Arizona, Tucson, AZ~85721}

\and

\author{Shoko Sakai}
\affil{Division of Astronomy and Astrophysics, University of California, 
Los Angeles, CA~90095-1562}

\and

\author{Edward J. Shaya}
\affil{University of Maryland, Astronomy Department, College Park, MD 20743}

\altaffiltext{1}{also Special Astrophysical Observatory of the Russian Academy of Sciences, 
 Nizhnij Arkhyz, 369167, Karachaevo-Cherkessia, Russia}
\altaffiltext{2}{Isaac Newton Institute of Chile, SAO Branch}
\altaffiltext{3}{Observatoire de Lyon, 9, avenue Charles Andr\'e, 69561, St-Genis Laval Cedex, France}

\begin{abstract}
Accurate distances to galaxies can be determined from the luminosities
of stars at the Tip of the Red Giant Branch (TRGB).  We use a
Maximum Likelihood algorithm to locate the TRGB in
galaxy color-magnitude diagrams. The algorithm is optimized by
introducing reliable photometric errors and a completeness characterization
determined with artificial star experiments. 
The program is extensively tested using Monte-Carlo simulations, artificial
galaxies, and a sample of nearby dwarf galaxies observed with HST/WFPC2
and ACS. Our procedure is shown to be reliable, to have good
accuracy, and to not introduce any systematic errors.
The methodology is especially useful in cases where the TRGB approaches
the photometric limit and/or the RGB is poorly populated.
\end{abstract}


\section{Introduction}
In the last decade, significant progress has been made in the study of the 
large-scale structure of the Universe.  Projects such as the Sloan Digital
Sky Survey are dramatically improving our knowledge of the kinematics, 
dynamics and structure in the distribution of galaxies at large distances.
At the same time, tremendous advances have been made on the small scale of
the local neighborhood as a result of searches for
nearby galaxies through wide-field optical
\citep{kk98,kk00,whiting2002}, and blind HI surveys \citep{meyer04}.
In the decade since the publication of a list of 226 galaxies 
with radial velocities less than 500~km~s$^{-1}$ \citep{k94},
the number of known nearby galaxies has doubled 
\citep{k04}. 
The Hubble Space Telescope (HST) and new large ground-based telescopes
make it possible to study stellar populations and measure accurate
distances to these nearby galaxies.

The luminosity of the tip of red giant branch (TRGB) provides a standard
candle that is being used to give accurate distances
to galaxies within 10~Mpc \citep{k05,m-a02}.
Over the last five years, members of our team have participated in
two snapshot surveys
of nearby galaxies using WFPC2 aboard HST, which have 
provided us with the material for distances of $\sim 10\%$ accuracy for 
about 150 nearby galaxies.
Further significant progress has been made with HST/ACS pointed observations.

The TRGB method is comparable in accuracy with the Cepheid Period-Luminosity
distance indicator \citep{s96,bel}. Furthermore, since
essentially all resolved galaxies are found to contain old stars,
the TRGB method is applicable to galaxies of
all morphological types, including spiral, elliptical and irregular galaxies.
The TRGB method also demands less telescope time to acquire a good distance
(only one HST/WFPC2 orbit is required for a TRGB detection in galaxies 
situated within 5~Mpc; HST/ACS gives a TRGB detection for galaxies out to 
10~Mpc in one orbit).
From early applications involving simply the visual identification of the
termination of the red giant branch, there has been a succession of more
sophisticated techniques to improve the
reliability and accuracy of the TRGB measurement \citep{lee,s96,mendez02}. 

About 200 galaxies have already been observed with HST/WFPC2 and 
HST/ACS, either by our collaboration or by others, with data 
suitable for a TRGB determination in the HST archive.
This outstanding material includes images of galaxies of essentially all
types, though dwarfs predominate.
The galaxies are mostly within 6 Mpc.
There are systems with high and low metallicities and mixes of stellar types.
Some targets have the TRGB situated well 
above the photometric limit but many have the TRGB only about 1 mag above 
the photometric limit.
The main aim of this work is to 
improve the TRGB estimation accuracy closer to the
photometric limit.
The existing sample gives us an excellent opportunity to compare TRGB detection
results for galaxies of different types and observed in different ways.
Ultimately, it is our intention to apply a consistent methodology to the
analysis of all the relevant data in the HST archive.

\section{Methodology}

Baade (\citeyear{baade44}) found that direct photographs of galaxies
in the Local Group showed the presence of a background sheet of red stars
and suggested that
the brightest stars in this sheet are like the brightest globular cluster
giant branch stars.
Sandage (\citeyear{sand71}) pointed out that
the brightest red stars in the underlying sheets of M~31, M~33, and IC~1613
have the same absolute magnitude, $M_V \approx -3.0\pm0.2$ mag.

According to modern stellar evolution theory, the tip of the first-ascent 
red giant branch marks the violent onset of core-helium burning in 
low-mass stars.
Observationally, this phenomenon causes a distinct and abrupt termination 
of the bright end of the red giant branch luminosity function. This 
discontinuity translates directly into an excellent distance indicator: 
the bolometric luminosity of the TRGB for low-mass stars is predicted to 
vary by only about 0.1 mag for ages ranging from 2 up to 15 Gyr 
\citep{ir83} 
and for metallicities encompassing the entire range represented by 
Galactic globular clusters of $-2.1 < [\rm{Fe/H}] < -0.7$ 
\citep{sc97}.
The discontinuity is also
found empirically to be stable at the $\sim 0.1$ mag level in the $I$ band 
for the same set of stellar properties.
A detailed review of the TRGB method is given by \citet{mf98}
and \citet{sakai99}.
Here we briefly describe the main features of the method 
and recent improvements.
 
Early applications of the TRGB method relied on a visual estimate
of the TRGB luminosity from a color-magnitude diagram (CMD).  
\citet{lee} defined the position of the TRGB in a reproducible and 
quantitative manner. They used a standard image-processing edge-detection (ED)
algorithm
employing the zero-sum Sobel kernel $[-2,0,+2]$ which, when convolved with the
stellar luminosity function, gives a maximum in its output at the 
luminosity where the count discontinuity is the greatest. The measuring errors
for the tip magnitude $I_{TRGB}$
were estimated to be typically 0.1--0.2 mag in their study.
The method as calibrated at that time is applicable over the range of 
metallicities $-2.2$ dex $< [\rm{Fe/H}] <-0.7$ dex.

A disadvantage of the binned approach of Lee et al. is that the TRGB solution 
depends both on the luminosity function (LF) bin size and bin placement.
\citet{s96} modified this method
for application with a smoothed, continuous LF.  The continuous $I$-band
LF is generated by replacing the discretely distributed stellar magnitudes
with the Gaussian-smoothed function:
\begin{equation}
\Phi(m) = \sum_{i=1}^N\frac{1}{\sqrt{2\pi}\sigma_i}\exp\left[-\frac{(m_i-m)^2}{2\sigma_i^2}\right],
\end{equation}
where $m_i$ and $\sigma_i$ are the magnitude and photometric error of the 
$i$th star and $N$ is the total number of stars in the sample. 
The ED filter is then defined by
\begin{equation}
E(m) = \Phi(I+\bar{\sigma}_m)-\Phi(I-\bar{\sigma}_m),
\end{equation}
where $\bar{\sigma}_m$ is the mean photometric error for all stars with 
magnitudes between $m-0.05$ and $m+0.05$.

The ED method can be very effective for galaxies when the TRGB
is located more than 2 mag above the magnitude limit of the photometry,
because in such cases the photometric errors at the tip are modest and 
crowding and incompleteness are not significant. 
The current estimate of uncertainties in TRGB measurements if well above the
photometric limit is 0.05 mag \citep{sakai04}.
However, the method becomes less
precise if sampling statistics are poor or incompleteness strongly affects
the LF within 1 mag of the TRGB location.  In such cases, the edge-detector
response becomes noisy even when the TRGB is clearly seen in the 
CMD.

As an alternative, a maximum-likelihood (ML) TRGB detection method 
was proposed by \citet{mendez02} (hereafter M02).
The ED algorithms find a maximum of the first derivative of
the luminosity function of the stars in the TRGB region, whereas in 
the ML analysis a predefined luminosity function is fitted
to the observed distribution of the stars.
The ED method has the advantage that it is model independent.
However the ED algorithm works by differentiation of the function 
determined by an ensemble of points which need to be smoothed. 
Smoothing with a small window gives a noisy result whereas
smoothing with a large window can introduce a bias.
The ML method does not have this disadvantage, but we need 
to know the
stellar luminosity distribution function and photometric errors distribution
for an adequate probability estimation. The ML method
proposed 
by M02 and a similar method by 
\citet{sakai04} do not
account for systematic errors introduced by crowding effects, Malmquist bias, 
and incompleteness.

Here we introduce the further development of the ML method
for the determination of the TRGB where we account for systematic photometric
errors. Artificial star experiments provide a reliable and well established 
method of photometric error estimation.  The HSTphot package
\citep{dol}
which we used for the photometry contains a powerful and 
flexible procedure for conducting artificial star tests.  Experiments
give us a measure of photometric errors from the differences between input 
and output magnitudes of artificial stars, and of the completeness function
from the input recovery rate.  The 
completeness, mean photometric error, and dispersion in error are determined
as a function of magnitude and affect the maximum likelihood luminosity 
function fit.

\section{Maximum Likelihood (ML) TRGB detection}    
The maximum likelihood method in our work is mathematically similar to the 
method introduced by M02. There are two main differences.
The first, already mentioned, is the use of a photometric errors function 
defined from artificial star experiments. 
It has been mentioned by many authors (see, for example, \citealt{gallart} and \citealt{dol02}) 
that artificial star tests are the only 
accurate way to solve the problem of photometric errors, blending, and 
incompleteness.  The procedure of artificial star tests involves the 
generation of a very large library
of artificial stars that cover the necessary range of magnitudes
and colors on the CMD so that the distribution of recovered photometry is 
adequately sampled.  The photometry program analyzes the generated artificial 
stars with the same routines
and parameter selections as for observed stars.
Photometric error estimations are determined by comparison
of predefined input magnitudes with photometered magnitudes.

The second main difference from M02 is the parameterization
of the RGB luminosity function.
For the ``theoretical'' luminosity function, we assume a simple power-law
with a cut-off in the TRGB region plus a power-law
of a second slope for a stellar population brighter than the TRGB.
\begin{equation}
\psi = \left\{
\begin{array}{ll}
10^{a(m-m_{TRGB})+b}, & m-m_{TRGB}\ge0 \\
10^{c(m-m_{TRGB})}, & m-m_{TRGB}<0
\label{e:lf}
\end{array}
\right.
\end{equation}
M02 use the slope $a=0.3$, which is also used in some 
other studies, but we allow some flexibility to the choice of the parameter 
$a$.  From our experience there are moderate variations around a slope 
$\sim 0.3$.  Filtering with color constraints can affect the slope.
The tests performed in this study do not confine the slope parameter $a$
but in the summary there will be a re-evaluation of this issue.

Definition of the photometric errors lead to the LF smoothing:
\begin{equation}
\varphi(m) = \int\psi(m')\rho(m')e(m|m')dm'
\end{equation}
where $\rho(m)$ is the completeness function and $e(m|m')$ is the error 
distribution function.  For the latter,
we assume a Gaussian distribution and take into account the bias in the 
photometric
error function; i.e. we use the first two moments of the distribution:
\begin{equation}
e(m|m') = \frac{1}{\sqrt{2\pi}\sigma(m')} \exp\left(-\frac{\left(m-\bar{m}(m')\right)^2}{2\sigma^2(m')}\right)
\end{equation}
Then the probability of a star detection of magnitude $m_i$ within interval $dm$ is:
\begin{equation}
P_i = \frac{\varphi(m_i|{\bf x})dm}{\int_{m_{\min}}^{m_{\max}}\varphi(m|{\bf x})dm},
\end{equation}
where $m_{\min}$ and $m_{\max}$ define the region of the validity of the 
parameterized distribution function and
{\bf x} is a vector of parameters.
Therefore, the probability of a given realization is $P=\prod_{i=1}^{N}P_i$. 
Let us define the minimization function as ${\cal L} = -\ln P$ and drop therms independent of model parameters, then:
\begin{equation}
{\cal L} = - \sum_{i=1}^{N}\ln\varphi(m_i|{\bf x}) + N\ln\int_{m_{\min}}^{m_{\max}}\varphi(m|{\bf x})dm
\end{equation}
Thus, in this method there are four parameters for the minimization, ${\bf x}(m_{TRGB}, a, b, c)$. 
The completeness function $\rho$ and photometric error function $e$ are 
defined from the artificial star tests.

\subsection{A Clean TRGB measurement}
The input data to our program includes observed and artificial star photometry 
results.  The observed photometry is extracted from HST/WFPC2 images with 
HSTphot \citep{dol} or from HST/ACS images with the modification
DOLPHOT \citep{dol05}.
From the artificial star tests we calculate statistics on mean deviations, 
dispersions, and completeness from the 
distribution of differences between 
recovered and  input $V$ and $I$ star magnitudes.
We use Gaussian weighting instead of binning
for a smoothed behavior of our functions. 
Our experience show that an 0.1--0.2 mag window size preserves the
original completeness behavior.

\begin{figure}
\centerline{\psfig{figure=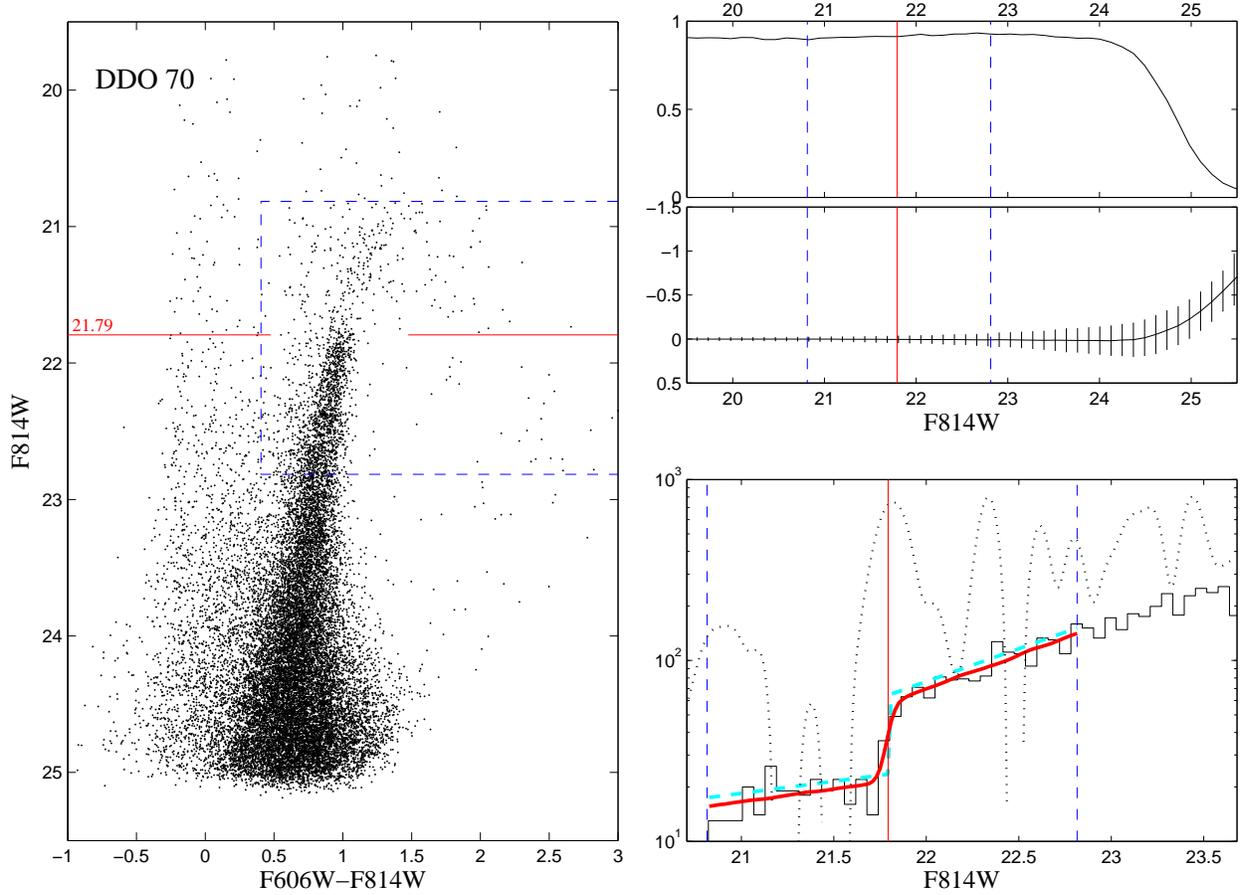,width=\textwidth,angle=270}}
\caption{\label{f:ddo70} The color-magnitude diagram (left panel) and the 
TRGB calculation results
for DDO~70. The upper right panel shows the completeness, photometric errors 
and dispersion in errors
(vertical bars) vs. the HST WFPC2 F814W filter magnitude.
The lower right panel gives a histogram of the F814W luminosity function.
The Gaussian smoothed LF first derivative is shown as a faint dotted line.
The resulting model LF convolved with photometric errors and incompleteness
is displayed as a bold solid line.  The intrinsic luminosity function
associated with the model is plotted as the dashed line with a
jump at the position of the TRGB. 
The fitted parameters are: m$_{TRGB}$ = 21.79 [21.76, 21.83], 
RGB slope a = 0.36 [0.30, 0.43],
RGB jump b = 0.44 [0.38, 0.50] and AGB slope c = 0.13 [0.03, 0.24].}
\end{figure}

As an example of our procedure, consider the color-magnitude diagram of 
DDO~70 shown in Fig.\ref{f:ddo70}.  Colors and magnitudes are in the HST
flight filters F814W, which transforms well to Cousins $I$, and F606W, which
transforms adequately to Johnson $V$. 
The most prominent feature of the CMD
is the RGB reaching more than 3~mag brighter than the photometric cutoff.
Stars immediately above the RGB are part of the asymptotic giant branch (AGB).
There is a modestly populated main sequence at F606W--F814W~$\sim 0$.
The peak of the first derivative of the Gaussian smoothed
luminosity function of stars provides a first guess of the location of the 
TRGB.
The dashed box in the CMD isolates the stars which were used in
the LF construction.
The magnitude limits are set $\pm 1$~mag brighter and fainter than the
preliminary estimated value of the TRGB.
The left color limit creates the restriction ${\rm F}606W - {\rm F}814W >0.4$.
These magnitude and color limits are
adjusted to suit the particular CMD.
The blue color limit is selected to eliminate main sequence stars.

There is no red color limit for a more subtle reason.  In instances
where the TRGB lies within a magnitude or so of the photometric limit,
the dashed box will encroach on this limit.  In fact there are two 
photometric limits
to consider because there are two filter bands.  There is the F814W magnitude
limit which is a horizontal cut on the CMD and there is the F606W magnitude
limit which causes a slanting cut on the CMD.  
If we insist that only stars with color information can contribute to the LF
then the slanting color cut creates a bias in the population of the LF.
The bias can be avoided if the LF is allowed to include stars clipped from 
the CMD because
they are very red; ie, stars detected in the F814W filter but too faint
to have been detected in the F606W filter.
Note that the point is moot in the case of DDO~70 because the analysis
is conducted at magnitudes well above the photometric limit.  This 
example demonstrates that the lack of a red limit has little effect on
the definition of the LF
because the number of stars redward of the RGB is small compared with the
number of stars in the RGB.

The current discussion does not give consideration to reddening, either
foreground or internal to the target galaxies.
This important topic will be considered in Paper~II of this series.

The output parameters -- TRGB value,
RGB jump, RGB slope and AGB slope -- and their uncertainties, are calculated 
from the analysis leading to the ML function fit.  Error intervals are 
68\% confidence level.  Results in the case of DDO~70 are
given in the caption to Fig.~\ref{f:ddo70}.
A detailed discussion of the method uncertainties is given in the next 
section.

\section{Tests}
\subsection{Recovering the TRGB with Monte-Carlo tests}
We performed extensive tests of internal precision through Monte-Carlo
simulations with known model parameters.
The model luminosity function was defined by 
random stars distributed according to Eq.~\ref{e:lf}.  Tests were run
with this function defined by 37, 371, 3706 and 37064 stars within 1~mag of 
the TRGB.
The photometric limit for our WFPC2 data is $I \sim 25.5$ mag and this limit 
defines the range of our tests.
In each case there were subtests with distance modulus varied such that 
$I_{TRGB}$ was shifted from 
22.0 mag to 25.5 mag in steps of 0.25 mag.  Each subtest contained 1000 
realizations.  The parameters $a$, $b$ and $c$ were the same for all
the tests: $a=0.3$, $b=0.3$ and $c=0.2$; quite similar to real galaxy LF
parameters.
We used the completeness function and photometric errors model from
real galaxy photometry (DDO 226: discussed later) to mimic real observations.
Such tests allowed us to study the influence of the number of observed stars 
and the problems that arise as the TRGB approaches the photometric limit. 

\begin{figure}
\centerline{\psfig{figure=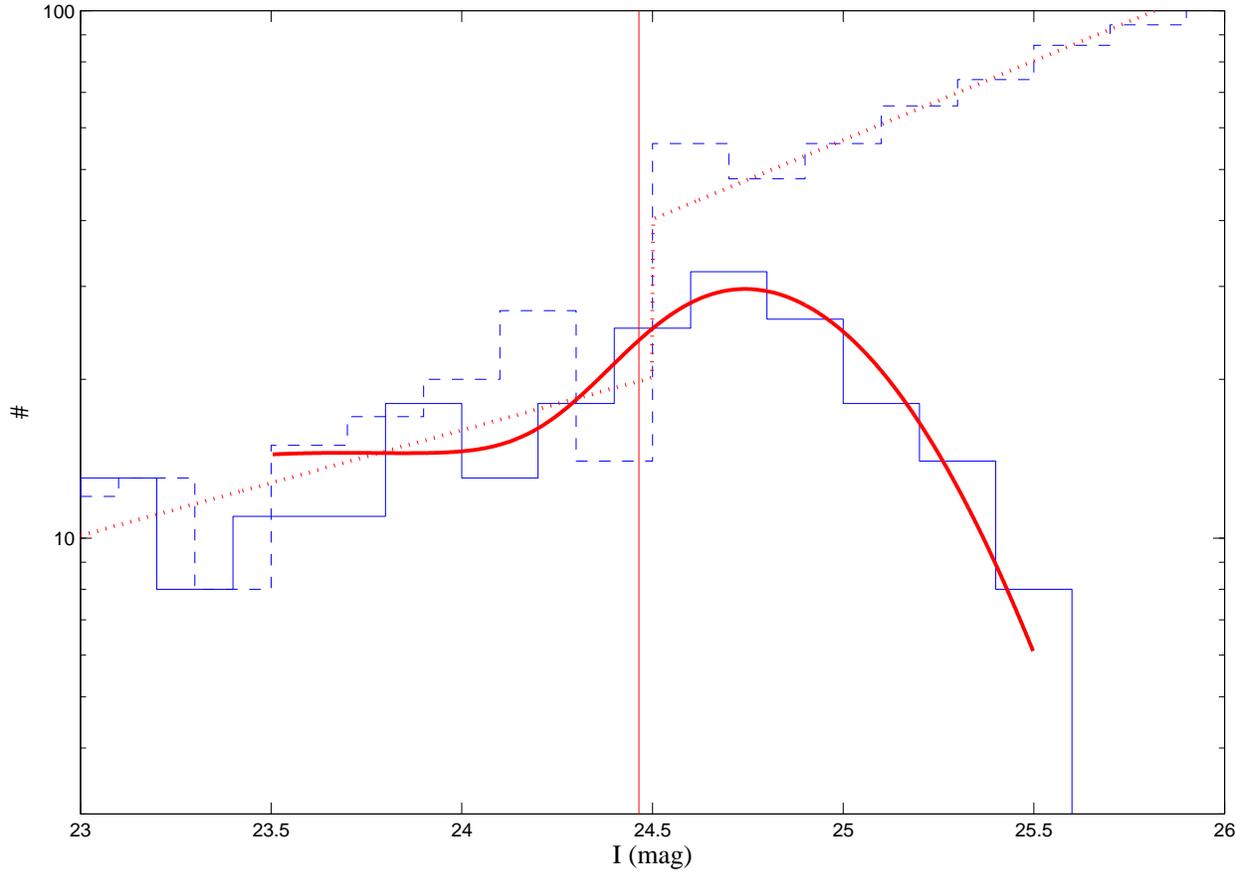,width=\textwidth,angle=270}} 
\caption{\label{f:tstfit} The luminosity function of an artificial star simulation. The input model
LF is shown with the dotted line, the dashed histogram is a randomly 
generated realization of the input LF, the solid histogram
is the LF after taking into account the observational errors, and the 
bold solid line represents
the result of the maximum likelihood analysis. The vertical line indicates 
the recovered TRGB position.}
\end{figure}

\begin{figure}
\centerline{\psfig{figure=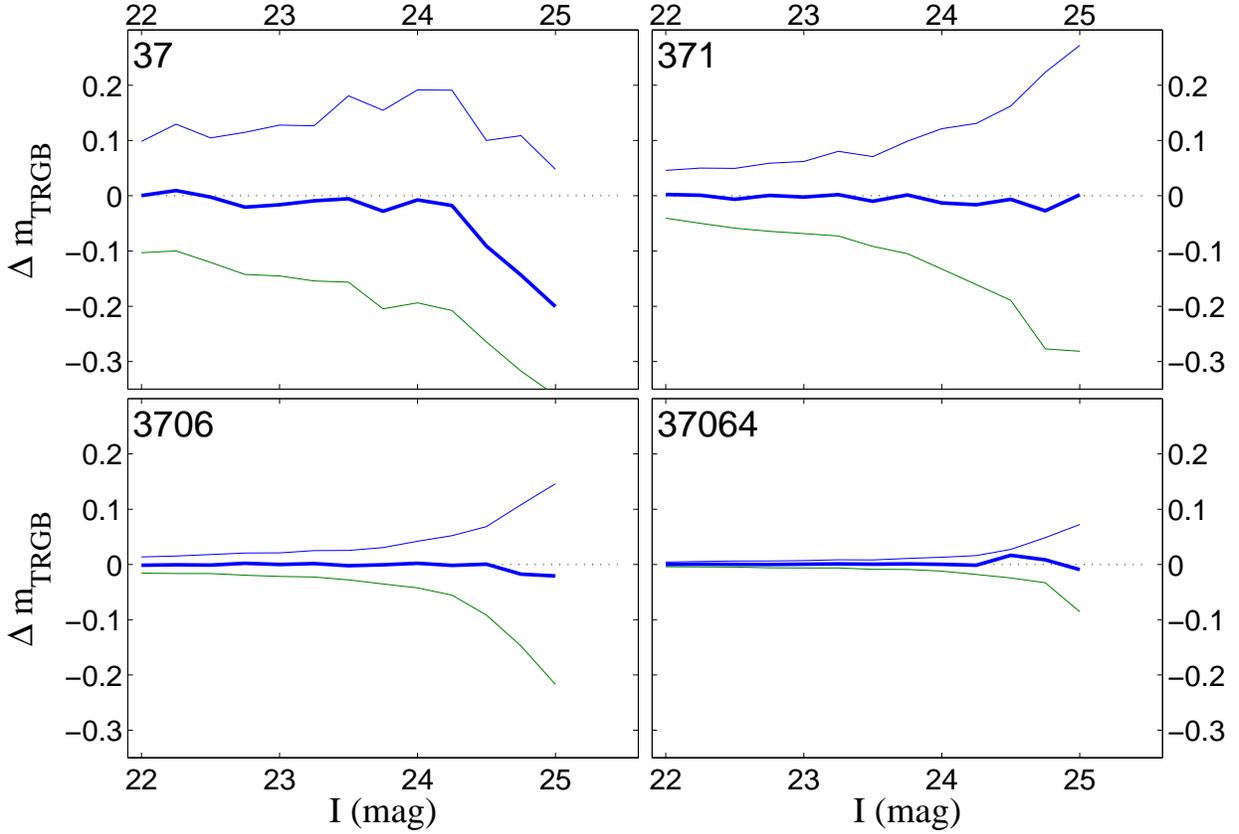,width=\textwidth,angle=270}} 
\caption{\label{f:tsttrgb} Median deviations of the recovered TRGB values 
minus input values are shown as bold lines.  There are 37, 371, 3706, and
37064 stars within 1 magnitude of the TRGB in the four respective panels.  
The thin lines show 75th percentile high and low deviations.
Experimental conditions reproduce snapshot WFPC2 observations.}
\end{figure}

\begin{figure}
\centerline{\psfig{figure=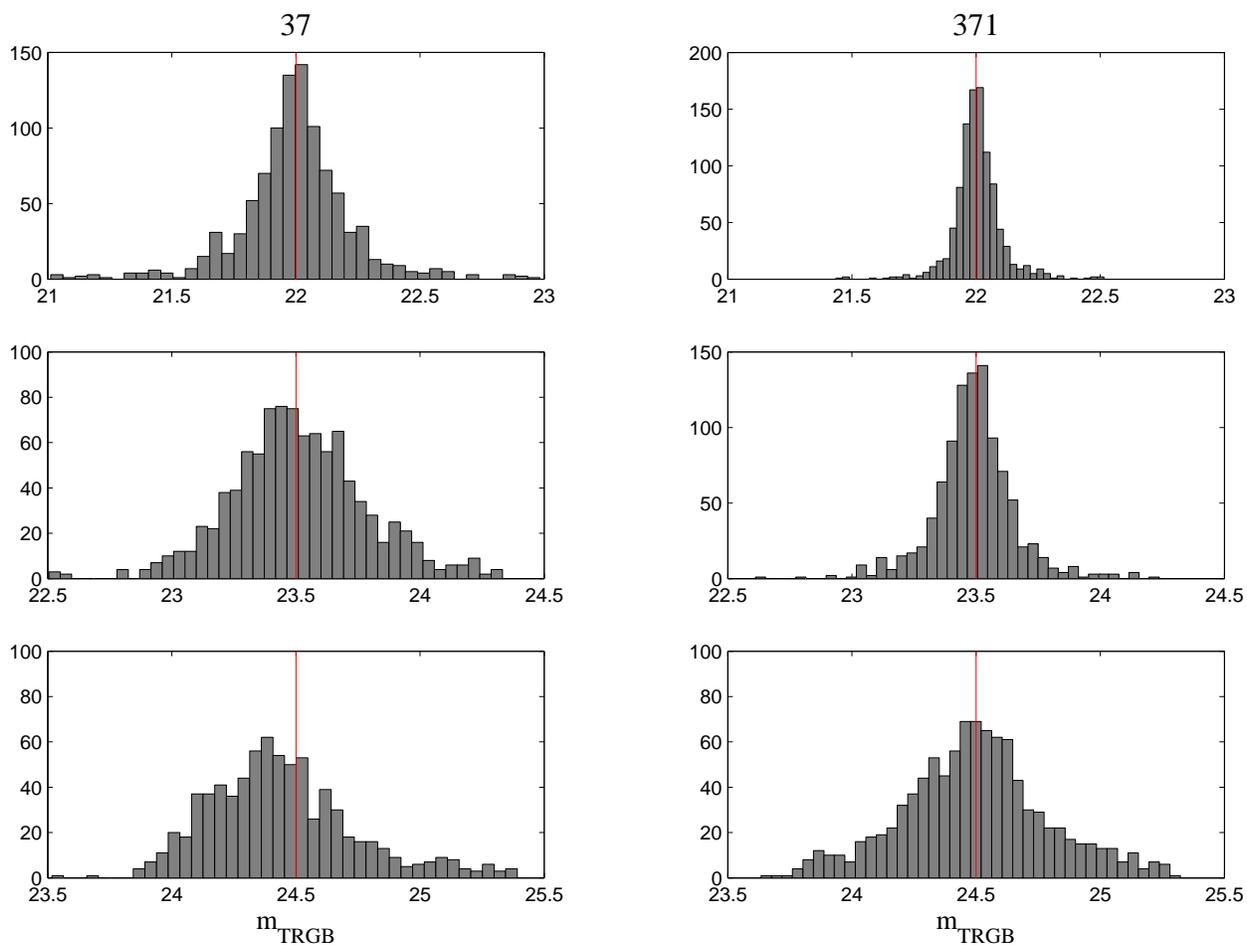,width=\textwidth,angle=270}} 
\caption{\label{f:tsthist} Distribution of reconstructed $I_{TRGB}$ for 
different sets of tests simulating snapshot WFPC2 conditions.
The left panel shows the results for a sample with 37 stars within 1~mag
of the TRGB and input $I_{TRGB}$ = 22, 23.5 and 24.5.
The right panel shows the results for a sample with 371 stars within 1~mag
of the TRGB with the same input $I_{TRGB}$.}
\end{figure}

\begin{figure}
\centerline{\psfig{figure=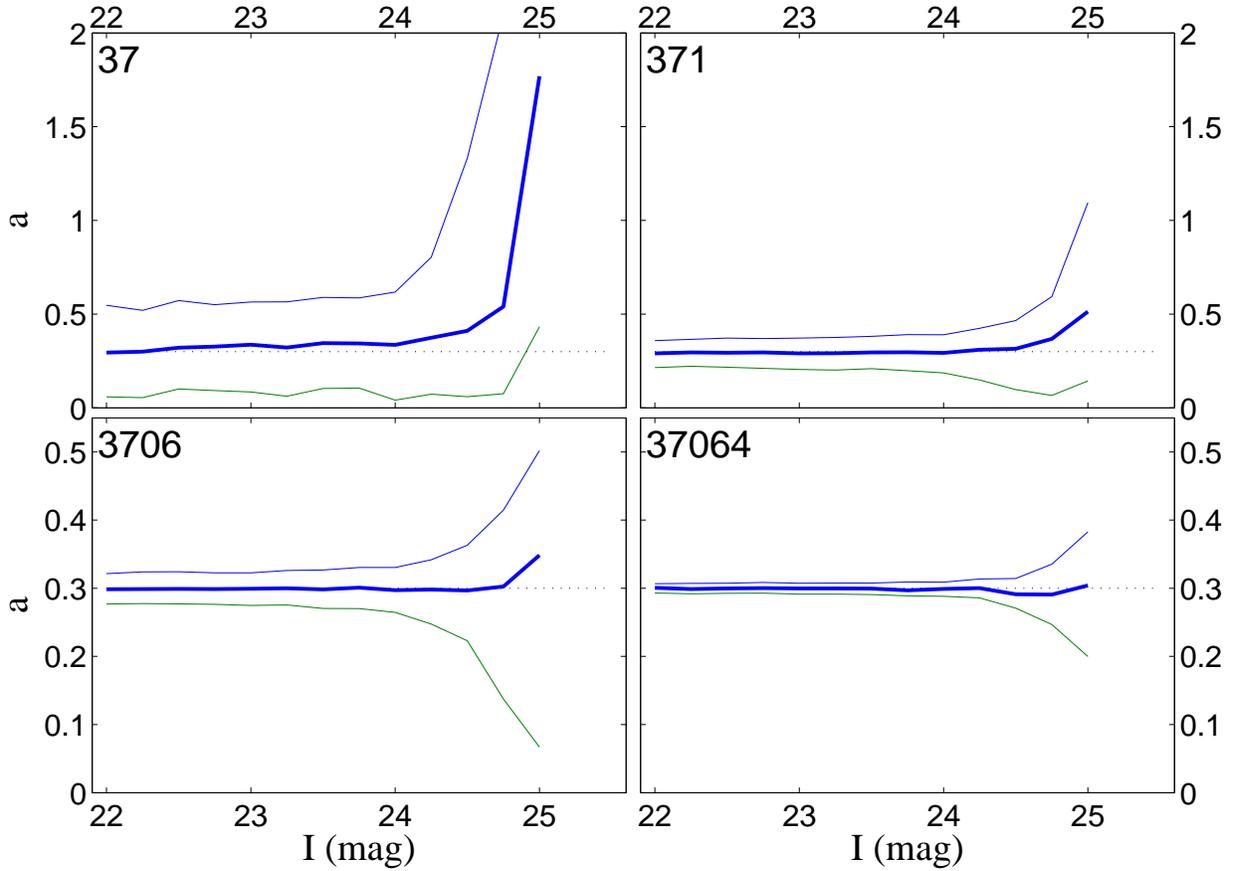,width=\textwidth,angle=270}} 
\caption{\label{f:tsta} Recovery of the RGB slope parameter $a$ with snapshot
WFPC2 conditions. The bold solid 
line is the median recovered value compared with the input value of 0.3. 
The thin lines show 75th percentiles of high and low deviations.}
\end{figure}

An example of a recovered LF is given in Fig.\ref{f:tstfit} where the input
TRGB is one magnitude above the photometric limit.
Median deviations of the recovered TRGB values from the input values are given
in Fig.\ref{f:tsttrgb}.  The four panels of the figure report results according
to the different number of the stars introduced into the LF.
The figure shows that our method does not introduce any bias even
when the TRGB is close to the photometric limit and it gives good accuracy even
in cases of small star samples. An exception is only the extremely
poor populated diagram where the TRGB is close to the photometric limit.
In Fig.\ref{f:tsthist} we show the distribution of reconstructed $I_{TRGB}$ for 6 different sets of tests.
It can be noted that the recovered $I_{TRGB}$ has a roughly Gaussian 
distribution except in
cases with very few stars or very close to the photometric limit.
The other three parameters of the model luminosity function are also 
recovered well, except in the limiting cases of few stars or near the 
photometric limit. As an example, Fig.\ref{f:tsta} shows the recovery of 
the RGB LF slope parameter.

\begin{figure}
\centerline{\psfig{figure=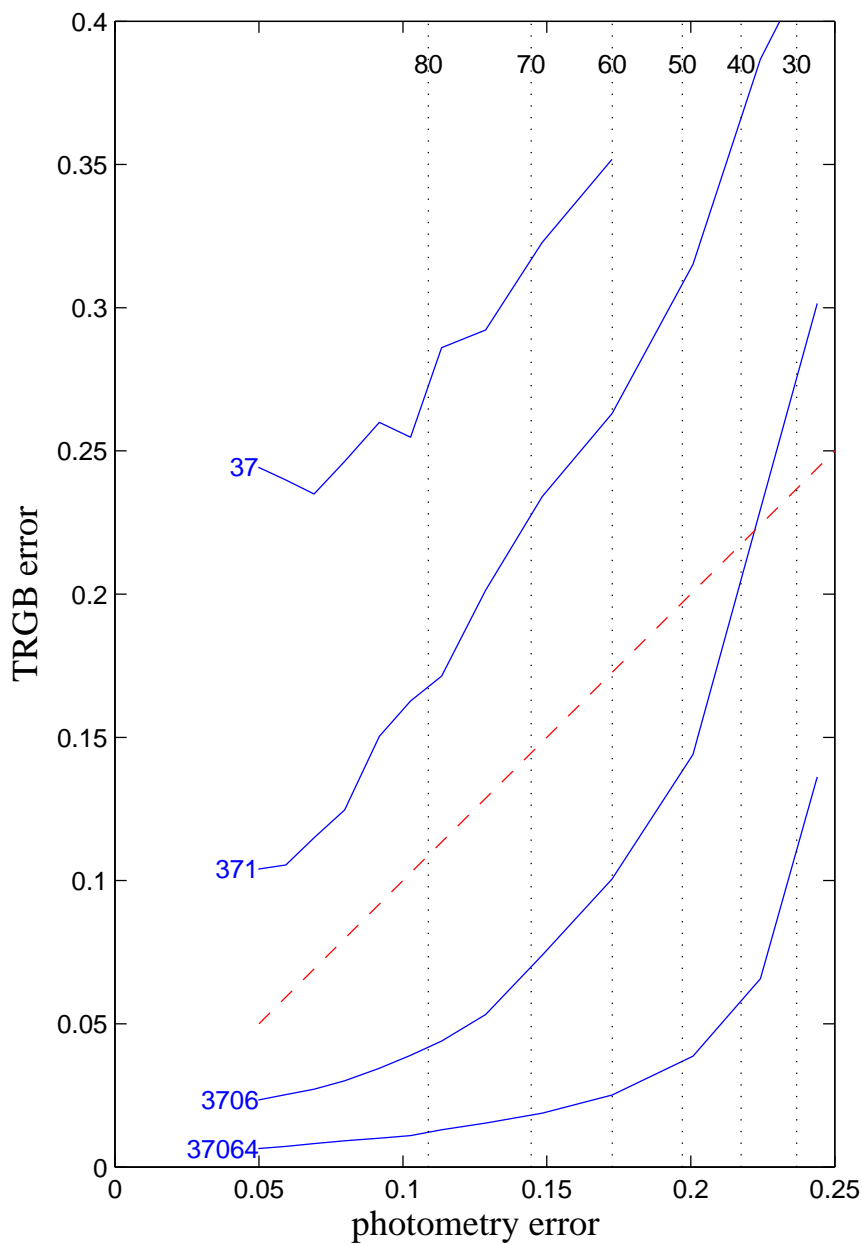,width=0.7\textwidth,angle=0}}
\caption{\label{f:errerr} TRGB uncertainties vs. photometric
error, received from artificial star experiments. The solid lines are mean TRGB uncertainties measured from
Monte-Carlo simulations. Each line is marked with the mean number
of artificial stars within 1~mag of the TRGB. The diagonal line
is the line of equal errors of TRGB measurement and stellar
photometric errors. The vertical dotted lines indicate the percent of
stars still participating in the analysis after taking 
observational effects into account.}
\end{figure}

Fig.\ref{f:errerr} shows the distribution of the measured TRGB uncertainty
vs. photometric errors of the individual stars for different numbers of
input stars. It can be seen from the figure that ML method accuracy
is strongly dependent on the number of stars.
The error of TRGB reconstruction becomes comparable to the photometric
error of individual stars if there are about 1000 stars in the analysis.
We can not expect an accuracy
better than 0.1 mag if there are less than
300 stars in the first magnitude below the TRGB.

\subsection{Comparison with other procedures using artificial galaxies}

To test our TRGB detection algorithm further, we intercompare some of 
the most commonly used methods and the ML procedure presented here on an additional 
set of artificial galaxy constructions that resemble real galaxies. 
We have applied the TRGB methods to  synthetic color-magnitude diagrams
(CMD's) constructed using the ZVAR code \citep{ber+02}. This code is based on the Padova stellar evolution tracks \citep{girardi2000}, and performs interpolation both in age and metallicity to produce a smooth distribution of stars.

As an aside, note that single-orbit ACS data extends $\sim 1.5$~mag deeper
in $I$ band than snapshot WFPC2 data, corresponding to equal sensitivity at a
factor $\sim 2$ greater distance.  Since we have both WFPC2 and ACS material
at our disposition, we explore tests tailored separately for the two cameras
and with our most common exposures times. 
The differences in sensitivity of the two cameras  are to be appreciated
in comparing real and simulated WFPC2 and ACS data.  A direct comparison
of WFPC2 and ACS data in a common field is given in Section 4.5.  The tests
to be discussed in this section are based on 
the photometric errors and the completeness effects
recovered using artificial star experiments based 
on real single-orbit observations obtained with HST/ACS.  

\begin{figure}
\centering
\resizebox{6.5cm}{!}{\includegraphics{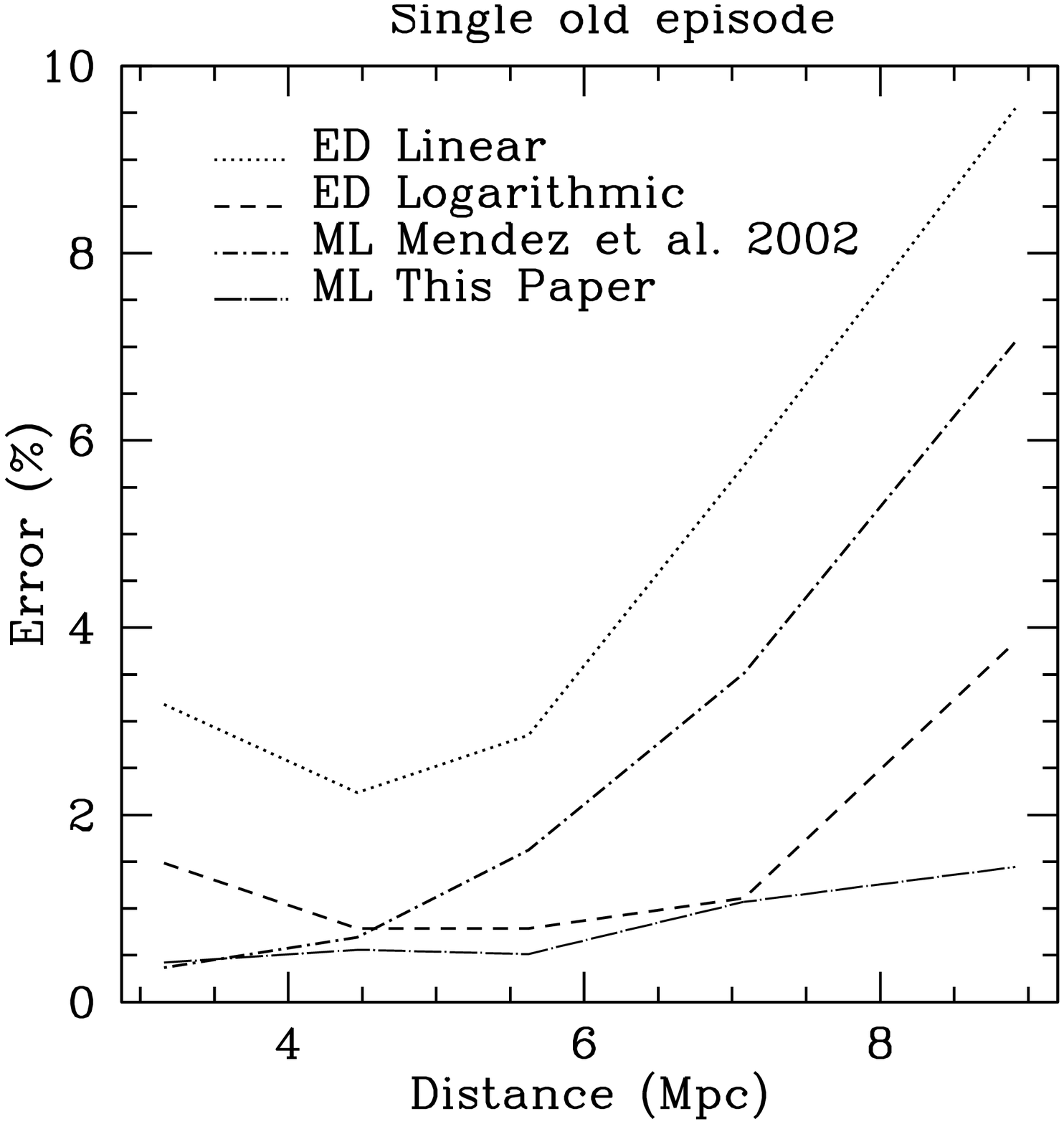} }
\resizebox{6.5cm}{!}{\includegraphics{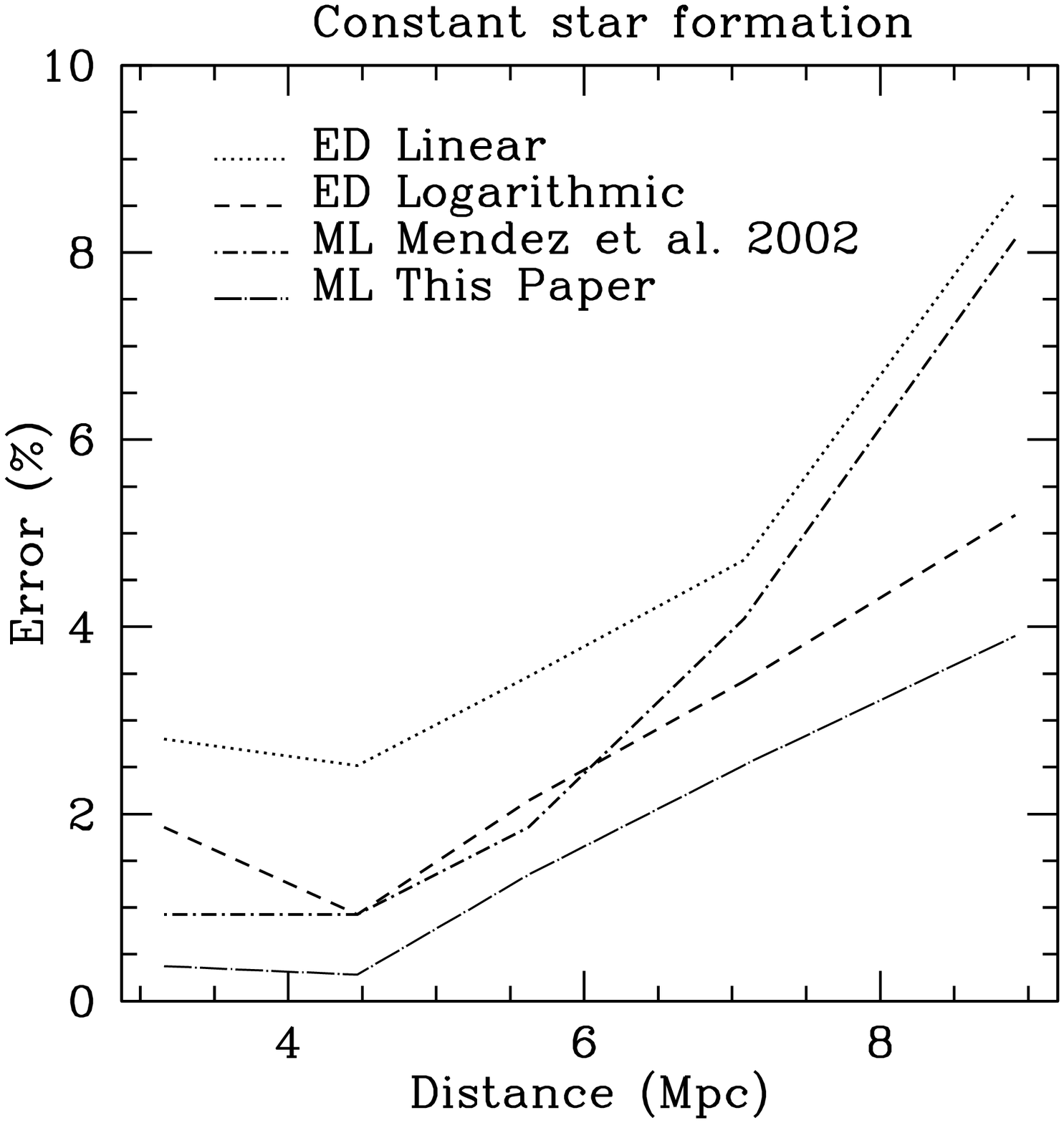} }
\resizebox{6.5cm}{!}{\includegraphics{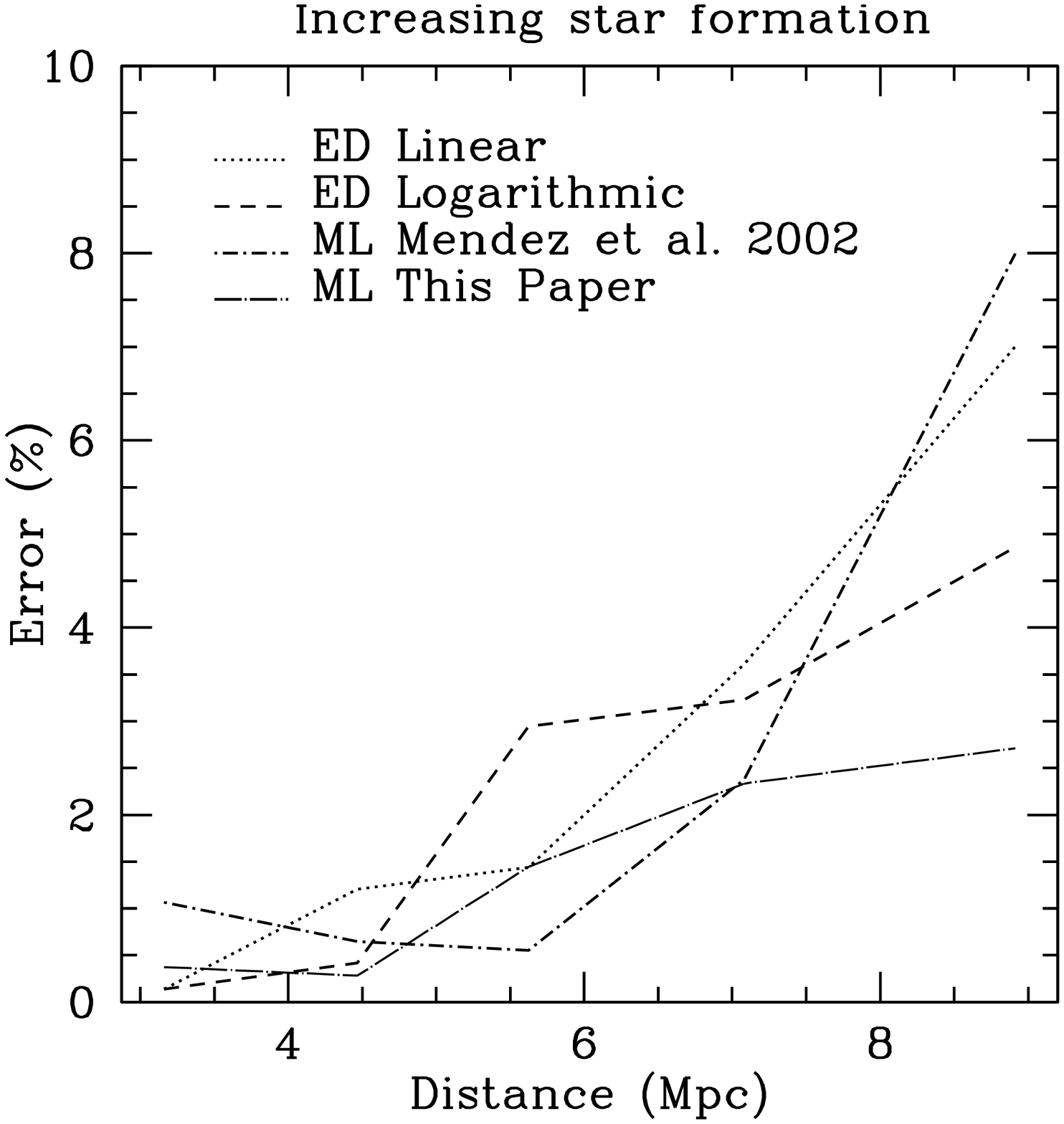} }
\caption[]{Contribution to the total distance error due to the TRGB 
detection alone. Top left panel: galaxies with a single old episode of
star formation. 
Top right panel: galaxies with continuous star formation.
Bottom panel: galaxies with an increasing star formation rate.
The tests assume the sensitivity of single-orbit observations with ACS.}
\label{fig:fig1}
\end{figure}

To avoid biases, no attempt at modeling the errors or the completeness
function was made, and the artificial stars were drawn from
a look-up table \citep[see][for details]{riz+02a}. Three different star formation histories were adopted, the
first consisting of an old single episode about 14 Gyr ago (closely
reproducing the case of a globular cluster), the second based on 
continuous star formation activity from 14 Gyr ago until now, and the third producing a mostly young galaxy. The
model galaxies were put at distance moduli ranging from 28 to 30, the
latter being the extreme case for a TRGB 
detection in a single orbit with HST/ACS. 
The TRGB detection methods we have tested are the ED (applied to both the 
linear and the logarithmic LF), and the ML 
\citep[both in the version of][and in the version presented in this paper]{mendez02}.
The results of our simulations are shown in Figure \ref{fig:fig1}.
The figure shows the contribution to the total distance error due 
to the TRGB detection alone.
A significant difference in the 
behavior of the methods we   
tested is apparent from these tests.  For the nearest cases all the methods 
give very good results. 
For more distant galaxies, ED methods seem to be more affected by 
problems connected with degraded photometric quality and by increased bias 
(stars preferentially scatter toward brighter magnitudes because the LF 
increases toward fainter magnitudes).

The right and bottom panels of Figures \ref{fig:fig1} show what happens 
for a  galaxy 
with a very strong AGB contamination, obtained by using a continuous star 
formation history, or a star formation that increases over time. In these 
cases, the RGB is superimposed on an extended and bright AGB component. 

Comparison of the upper left panel of Figure \ref{fig:fig1} with the other
two panels
clearly shows that all methods are basically insensitive to the presence of 
AGB stars, and that ML methods produce significantly smaller errors at all 
distances and for all the combinations of star formation histories.

\subsection{Poorly populated CM diagrams}

\begin{figure}
\centering
\resizebox{6.5cm}{!}{\includegraphics{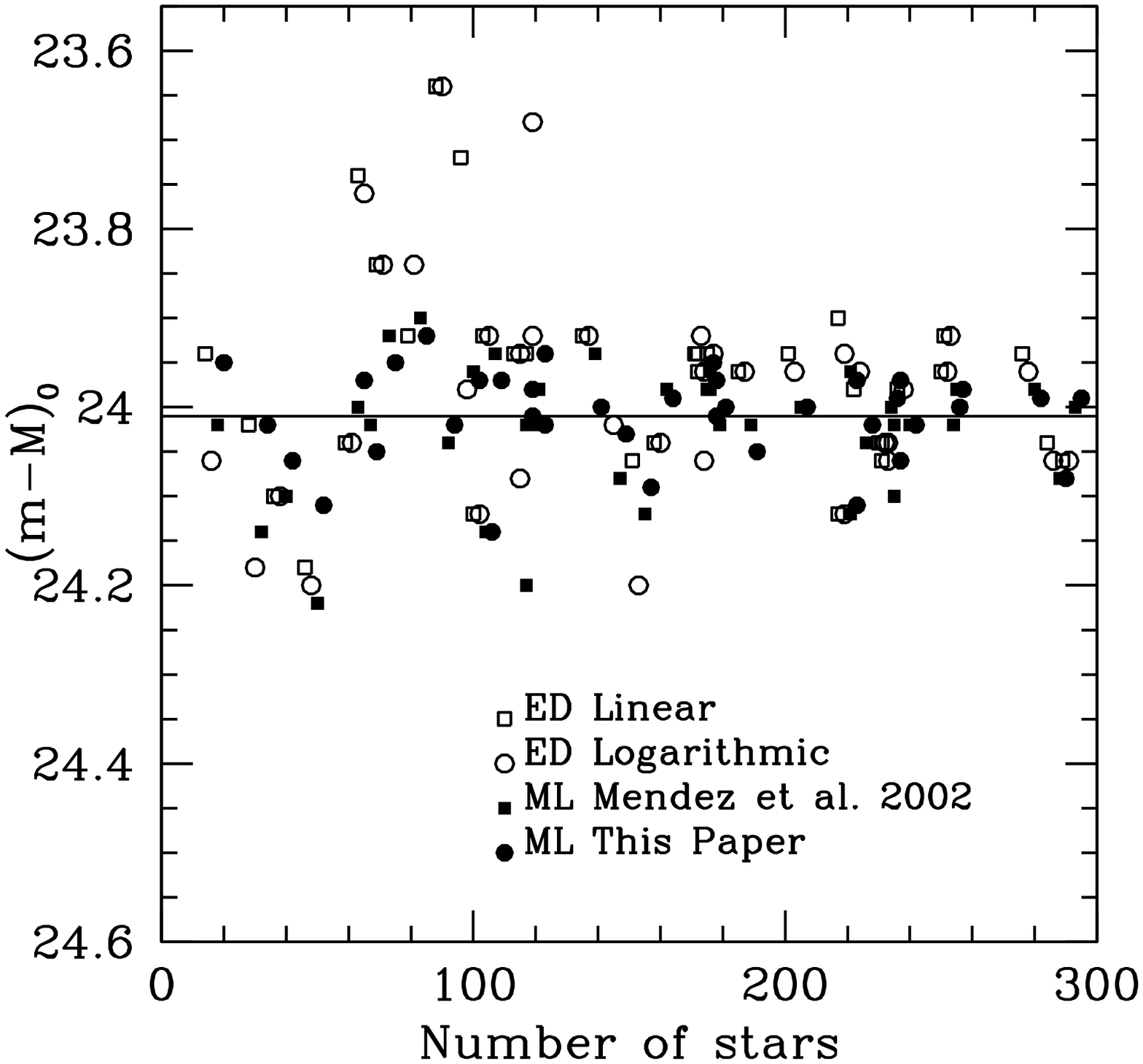} }
\resizebox{6.5cm}{!}{\includegraphics{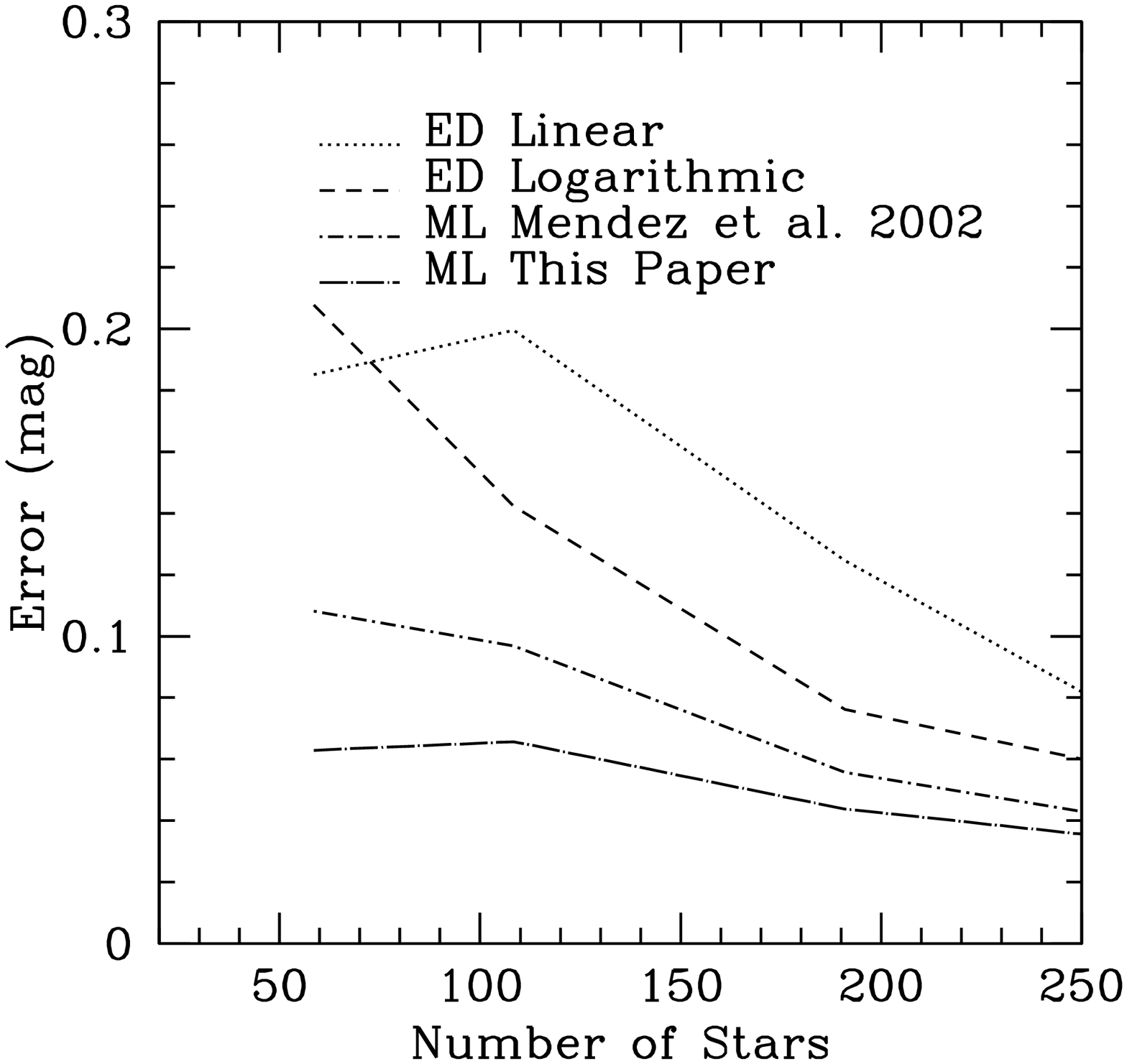} }
\caption{\label{fig:fig4}TRGB detection against the number of stars in the 
first magnitude bin. Left panel: deviations from the expected TRGB position 
at 24.01. Right panel: r.m.s. of the different methods against number of 
stars in the first magnitude bin.}
\end{figure}

To consider the effect of poorly populated CM diagrams in greater detail, 
we investigated 
the case of a globular cluster-like galaxy, at an intermediate distance of 
$(m-M)_0=28$. The star population in the CM
diagram is quantified by the number of stars in the first magnitude bin 
after the tip 
(hereafter, $N^{(-1)}$). We generated galaxies with $N^{(-1)}$ between 300 
and 30  and applied the detection methods to all cases. Results are shown 
in Figure \ref{fig:fig4}.

The left panel of Figure \ref{fig:fig4} shows that there is a general trend 
for a large scatter in the TRGB detection when $N^{(-1)}$ falls below 100. 
This result is consistent with previous studies \citep{fre+mad95}. 
Careful scrutiny of this plot also shows that the average dispersion of 
measurements is significantly lower at any $N^{(-1)}$ for ML methods, 
compared to ED methods. This point is further demonstrated by the right panel 
of Figure \ref{fig:fig4}, which shows the r.m.s. of TRGB detections against 
$N^{(-1)}$. Not only is the r.m.s. of ED methods always higher, but 
ED methods tend to break down at a $N^{(-1)}$ of about 100 stars, while 
ML methods seem to be able to produce reasonable results at least down to 
50 stars.  Incidentally, we note that we do not find in our simulations 
the large systematic deviations pointed out by \citet{fre+mad95}. The 
difference is most likely due to the presence in our simulations of AGB stars.
\citet{fre+mad95} use fiducial lines from globular clusters populated with 
power-law distributions, so they have no stars brighter than the TRGB. 
Removing stars from their simulated galaxies can only produce fainter TRGB 
detections, while our use of synthetic stellar populations produced with 
stellar models results in deviations in both directions.

In conclusion, we verify that ML appears to perform better than ED methods 
especially when the level of the TRGB approaches the detection limit. 
We also verify that even a prominent AGB 
component does not significantly affect the detection. Finally, we applied 
the methods to poorly populated diagrams, and found that ED methods produce 
significantly higher errors for each level of $N^{(-1)}$. ML methods seem to 
be able to detect the TRGB with as few stars as $N^{(-1)}=50$.

\subsection{Comparison of ML and ED procedures using real data}

To estimate the external uncertainties of our method we compared the results
of ML TRGB detection with the results of the modified 
ED algorithm \citep{s96} for the same objects.  
For the comparison we selected 10 dwarf galaxies from the snapshot 
surveys (HST programs 8601 and 8192). We have taken 
both dIrr and dSph dwarfs at different distances, extending to 5~Mpc.
The stellar photometry was done with the software HSTphot \citep{dol} in 
the course of the snapshot programs with artificial star photometry added.
The WLM galaxy (DDO~221) 
was included in the sample to provide a very nearby case with a deep
CMD. The HST/WFPC2 observation of DDO~221 were made within HST programs 6898 
and 6813, and photometry results along with the results of artificial star 
experiments were
taken from J. Holtzman's HST Local Group stellar photometry archive
\citep[][http://astronomy.nmsu.edu/holtz/archival/html/lg.html]{holt}.

\begin{figure}
\centerline{\psfig{figure=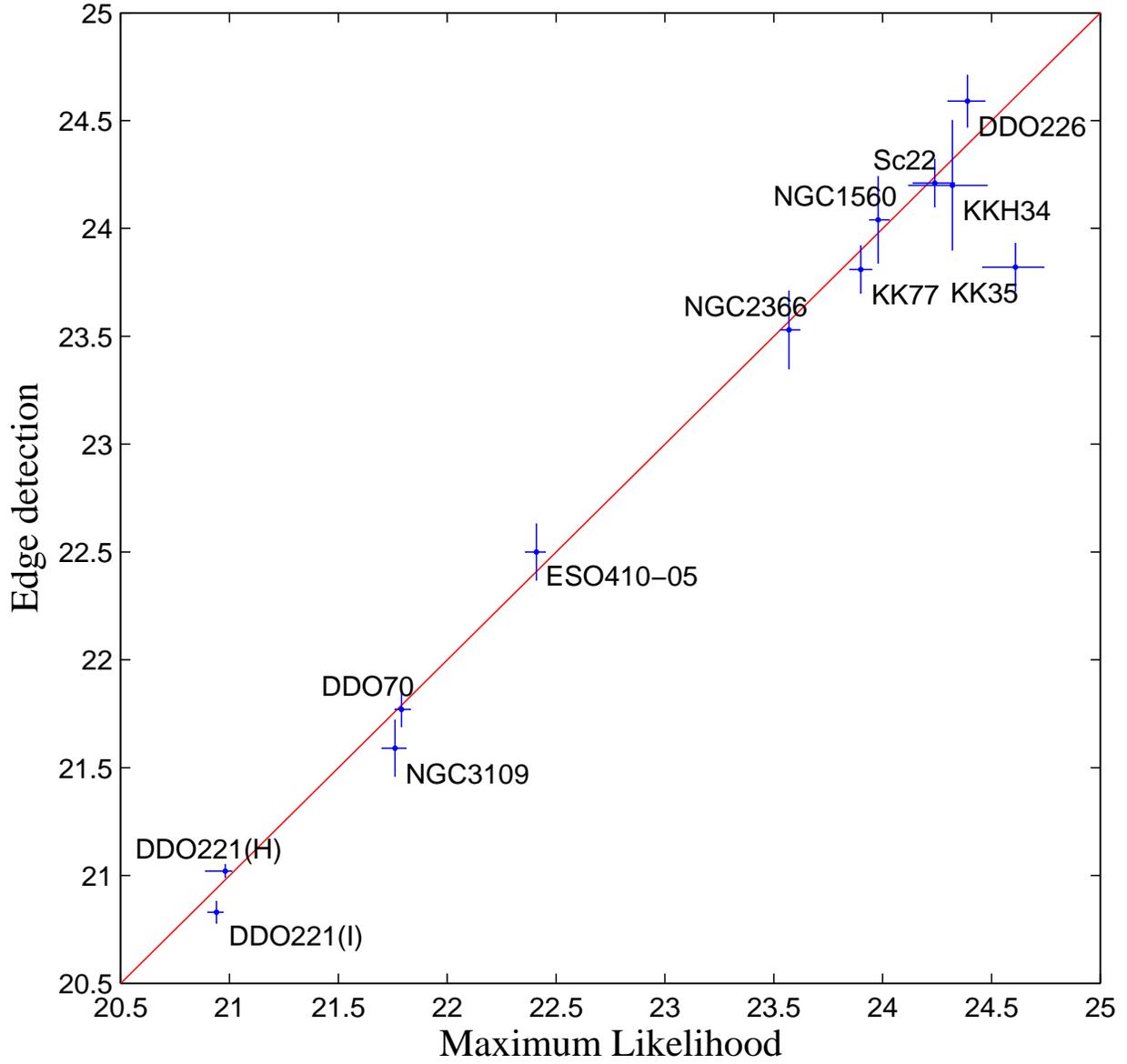,width=\textwidth,angle=0}}
\caption{\label{f:edml} Comparison of TRGB detection results obtained
alternatively with the 
maximum likelihood and edge detection algorithms.}
\end{figure}

The results of the comparison of the two methods are presented in 
Fig.\ref{f:edml}.  The figure shows $I_{TRGB}$ 
measured with the ML method vs. $I_{TRGB}$ 
measured with ED method. The horizontal error bars are the 
ML one-sigma errors and the vertical bars are the equivalent
errors with the ED method.

The agreement of the two measurements for two fields in DDO~221 are
good. The mean difference between two methods is 0.07 mag, whereas the 
difference between the two fields measured with the same method is 0.05 for 
the ML
and 0.19 for the ED method.  
For many of the galaxies the TRGB is quite close to the photometric limit 
which challenges both methods. A marked disagreement between
the two methods can be seen in the case of KK~35 where it is likely that the
ED measurement is picking up the onset of the AGB.
The maximum likelihood estimates are expected to be more effective in cases 
where the TRGB lies close to the photometric limit.
It should be noted that the photometric error bars are smaller for the 
ML method in most cases.
The maximum likelihood procedure has
an advantage due to the incorporation of reliable photometric errors and 
completeness function through the  
artificial star experiments.
We did not find any correlation of $I_{TRGB}$ (maximum likelihood) and 
$I_{TRGB}$ (ED) difference with the 
morphological type or absolute magnitude of the sample galaxies. 

In general, the comparison of the maximum likelihood and ED 
methods shows
that both methods work well, with comparable accuracy, when the RGB is 
well-populated and is situated at least 2 mag above the photometric limit. 
However, when the
TRGB is closer to the photometric limit and/or poorly populated, the 
ED measurement can give a result with poor accuracy and bias.
On the other hand, tests suggest that
the ML method, combined with a proper description of 
incompleteness and real
photometric errors, gives reliable results whenever the TRGB is accessible. 

\subsection{A comparison between two fields in the same galaxy}

It is interesting to compare the ML TRGB detection 
results when more than one field of a galaxy was observed, and also when there
are several  exposures within the same field.  The comparisons allow us 
to estimate the internal accuracy of the ML method, 
the influence of signal-to-noise on the detection results, and
the possible effects of population differences.
For the comparison we selected the dIrr galaxy NGC~2366, for which there is
an impressive quantity of HST/WFPC2
observational material in the HST archive. Two samples are appropriate for our
purposes: a sequence of
exposures with the F555W and F814W filters within program 8769 and
pairs of exposures of a separate field in the F606W and F814W filters within 
program 9318.  Details about exposures and the result of our
measurements are given in Table~1.
 
\begin{figure}
\centerline{\psfig{figure=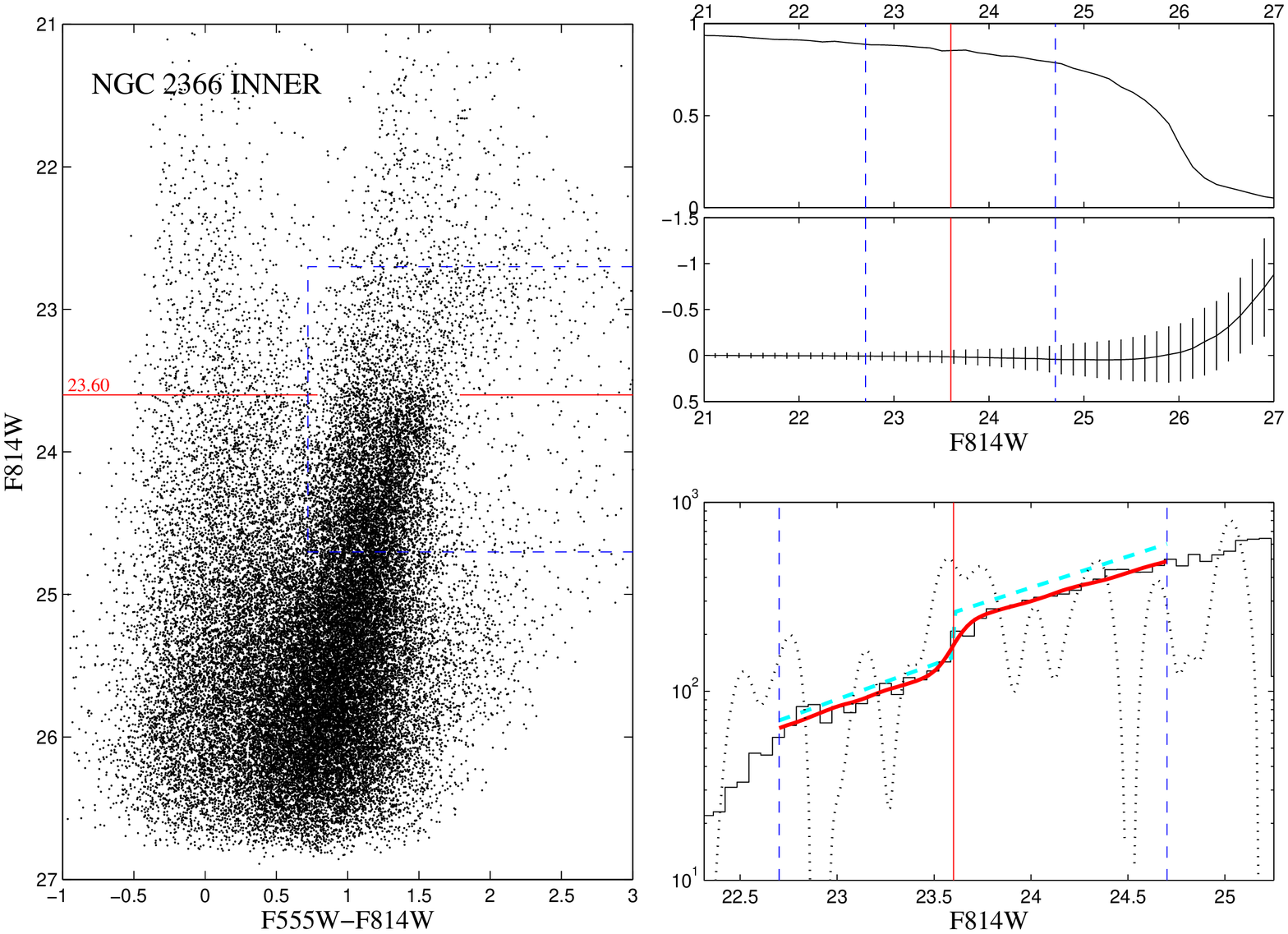,width=0.8\textwidth,angle=270}} 
\centerline{\psfig{figure=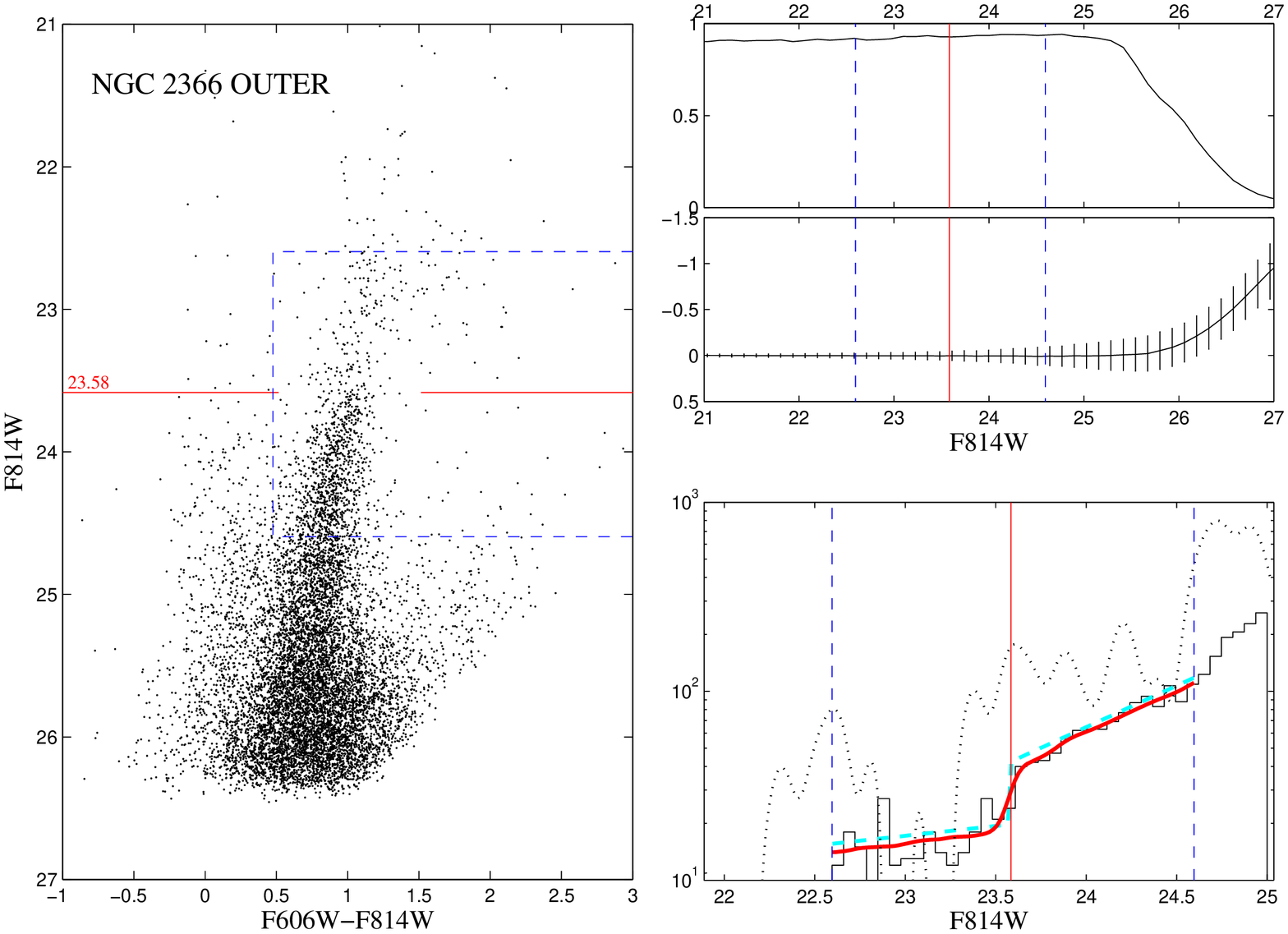,width=0.8\textwidth,angle=270}} 
\caption{\label{f:n2366} The TRGB detection result for the inner field 
of NGC~2366 (upper panel) and outer field of NGC~2366 (lower panel).  
The F814W vs. F555W--F814W or F606W--F814W (equivalent I,V--I) 
color-magnitude diagrams are 
at the left. 
Completeness, error, and error dispersion functions are at top-right.
At lower-right, the stepped line
is the histogram of the observed LF and the bold line is the LF
accounting for uncertainties and incompletion.}
\label{fig:n2366}
\end{figure}

The agreement of TRGB values
in both fields is excellent; the difference is 0.02 mag. 
The first field covers the inner part of NGC~2366.
The color-magnitude diagram of the sum of all images is presented 
in the top panel of Fig.\ref{f:n2366}.
The TRGB is situated well above the photometric limit in this case, and it is 
easily detected
with our procedure. 
The color-magnitude diagram has  huge blue and red supergiant populations, 
as well as rich RGB and AGB populations.
The comparison of results between the summed photometry
and the single exposure photometry in this inner field again demonstrates the 
internal accuracy is good: the mean deviation is 0.02 mag. The consequence
of better signal with the summed data is negligible because the TRGB is 
already about 2 mag above the photometric limit with the individual exposures.
The photometric results for the outer field of NGC~2366 are presented in 
the lower panel of Fig.\ref{f:n2366}. There is only a small contamination 
here from red 
and blue supergiant stars,
yet the AGB is well-populated in this outer region of the galaxy.  
The good agreement of the TRGB fits in the two fields demonstrates that the 
presence of a very prominent young population does not affect the TRGB 
measurement.

\subsection{Comparison of WFPC2 and ACS observations of the same galaxies}

\begin{figure}
\centerline{\psfig{figure=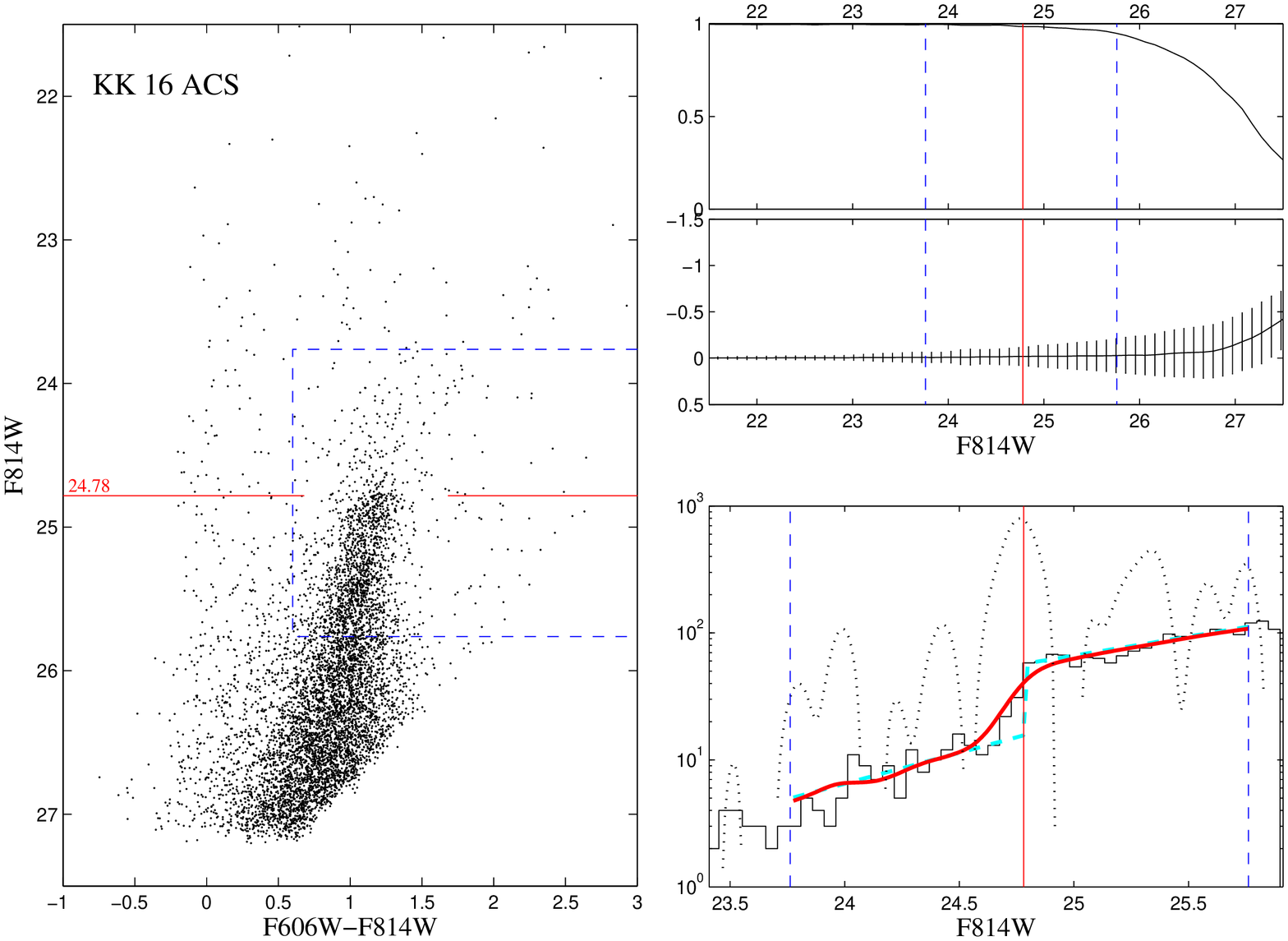,width=0.8\textwidth,angle=270}}
\centerline{\psfig{figure=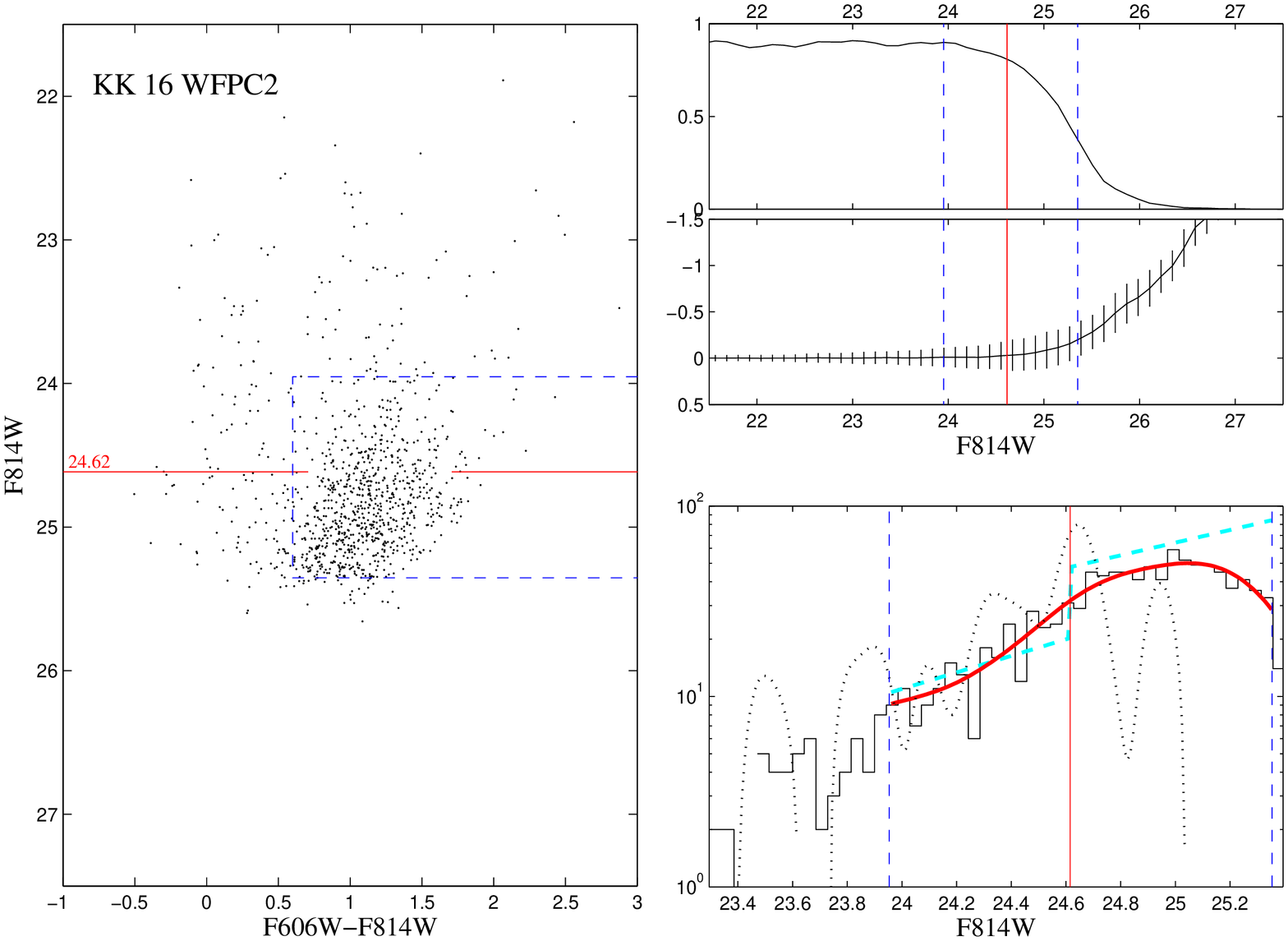,width=0.8\textwidth,angle=270}}
\caption{\label{f:acswfpc} The color-magnitude diagram and the TRGB
calculation results for KK~16 ACS observations (upper panel)
and the same for KK~16 WFPC2 observations (lower panel).}
\end{figure}

With the new instrument Advanced Camera for Surveys (ACS) on
HST, the capability to measure galaxy distances with the TRGB method
has improved significantly.  ACS provides a wider field of view, better 
resolution, and increased sensitivity.  We can now obtain accurate
galaxy distances at 10 Mpc
with one orbit of HST observing time.  An example of the enhanced
capability is provided by the dwarf galaxy
KK~16 which was observed both with WFPC2 (program 8601)
and ACS (program 10210). In the case of the WFPC2 observations
(exp = 600s in both F606W and F814W),
we find F814W$_{TRGB}$ = 24.62$\pm0.10$ (D = 5.0 Mpc) (lower panel 
Fig. \ref{f:acswfpc}).
Definitely, this TRGB detection is near the limit for a WFPC2 snapshot 
observations.
The ML procedure picks up the TRGB near the photometric limit where
the photometric errors and bias are substantial.  By contrast,
the observations of the same galaxy with ACS (and slightly longer
exposures of 1226s in F814W and 934s in F606W) give a
very reliable CMD sampling with F814W$_{TRGB}$ = 24.78$\pm$0.04 (D = 5.7 Mpc) 
(upper panel Fig. \ref{f:acswfpc}).
The WFPC2 snapshot exposures only detect the onset
of the RGB for KK~16, and the distance is underestimated. However, even
in this extreme case the procedure does not fail egregiously.

\begin{figure}
\centerline{\psfig{figure=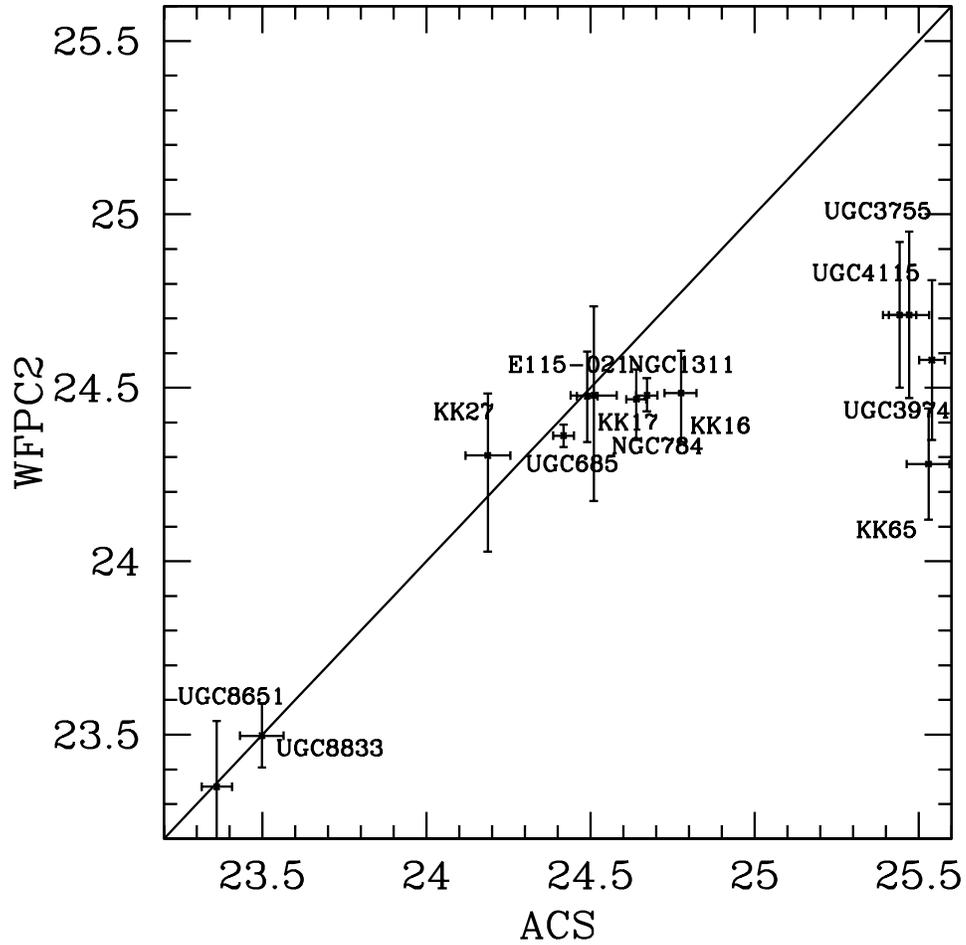,width=0.8\textwidth}}
\caption{\label{fig:wfpc2-acs}
Comparison of single orbit ACS and WFPC2 TRGB determinations.
Single orbit WFPC2 measures are unreliable at $I_{TRGB} > 24.5$.
It is expected that single orbit ACS measures become unreliable at
$I_{TRGB} > 26$.
}
\end{figure}

There are circumstances that can give rise to wildly erroneous results.
The problem is made clear in Figure \ref{fig:wfpc2-acs} which provides 
a further comparison between WFPC2 and ACS observations.  Measurements
in good agreement lie along the 45 degree line.  If the TRGB is bright
then there is excellent agreement, but fainter than $I_{TRGB} \sim 24.5$
the WFPC2 values fall below the line.

With single-orbit WFPC2 observations, the photometric limit is $I \sim 25.5$.
In the case of KK~16 that was just discussed, the TRGB is being identified 
but the location is biased by it's proximity to the photometric cutoff.
On the other hand, in the cases of the four galaxies at the extreme right
of the figure it is demonstrated by the ACS observations that 
the TRGB is below the WFPC2 detection limit. The WFPC2 measurements
that have been reported for these galaxies \citep{kar03}
are all based on the detection of the onset of the AGB.

We have not found an algorithmic way to ensure that the TRGB is being
isolated and not the brightest AGB stars when the feature is near the 
photometric limit.  The ratio of red stars to blue stars (cutting in color 
at $V-I = 0.6$) in the first magnitude below the putative tip can give a
good clue.  AGB stars are usually (but not always) accompanied by 
substantial numbers of blue main sequence stars at comparable $I$ magnitudes.
A sign of a dominant old population RGB appearing at the faint limit is a
ratio of red to blue stars greater than 3.5.  However, the only convincing
way to establish if one is seeing the TRGB is to see the AGB above it.
The onset of the AGB is $\sim 1$~mag brighter than the TRGB so should be 
evident if there is a chance to measure the TRGB.

The problems seen in the WFPC2 data will obviously also arise in the ACS
data but, with single orbit observations, is displaced $\sim 1.5$~mag
fainter.  Problems will begin at $I_{TRGB} \sim 26$ due to the faintness
limit at $\sim 27$.  For the moment, there are hundreds of galaxies still
to be observed with $I_{TRGB} < 26$. 

\subsection{Example of a difficult case}

\begin{figure}
\centerline{\psfig{figure=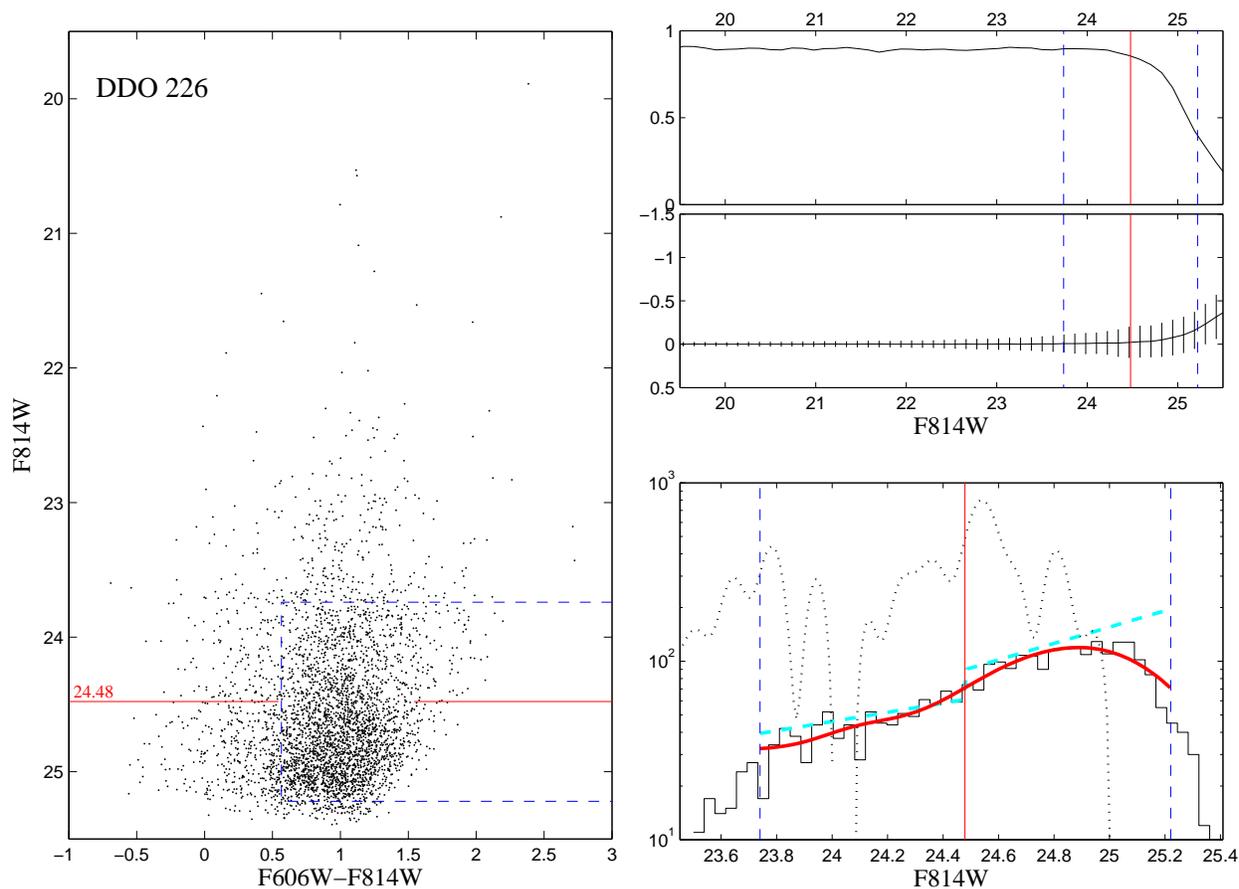,width=\textwidth,angle=270}} 
\caption{\label{f:ddo226} The color-magnitude diagram and the TRGB 
calculation results
for DDO~226. The upper right panel demonstrates the completeness, photometric 
errors, and dispersion 
(vertical bars) vs. F814W mag. The lower right panel is a histogram of the 
F814W luminosity function
(solid line), Gaussian smoothed LF first derivative (dotted line), 
the model LF accounting for errors and incompletion (bold solid line),
and the corresponding intrinsic LF (dashed line with step at the TRGB).}
\end{figure}

The color-magnitude diagram of DDO~226 is demonstrated in Fig.\ref{f:ddo226}.
Exposures of 600 sec were obtained in both F606W and F814W filters.
This dwarf irregular galaxy has a substantial AGB population, which is
clearly seen in the CMD, and a number of blue and red supergiant stars.
The red giant branch and AGB are widened by errors because they are close
to the photometric limit. 
The TRGB is situated about one
magnitude above the photometric limit. In this case the influence of photometric
errors and incompleteness makes the TRGB detection especially difficult. However,
using the photometric error modeling from the artificial star experiments
still allows us to obtain reliable results:
F814W$_{TRGB}$ = 24.39$\pm$0.08 at the 68\% confidence level.

\subsection{Spatial selection in a difficult case}

\begin{figure}
\centerline{\psfig{figure=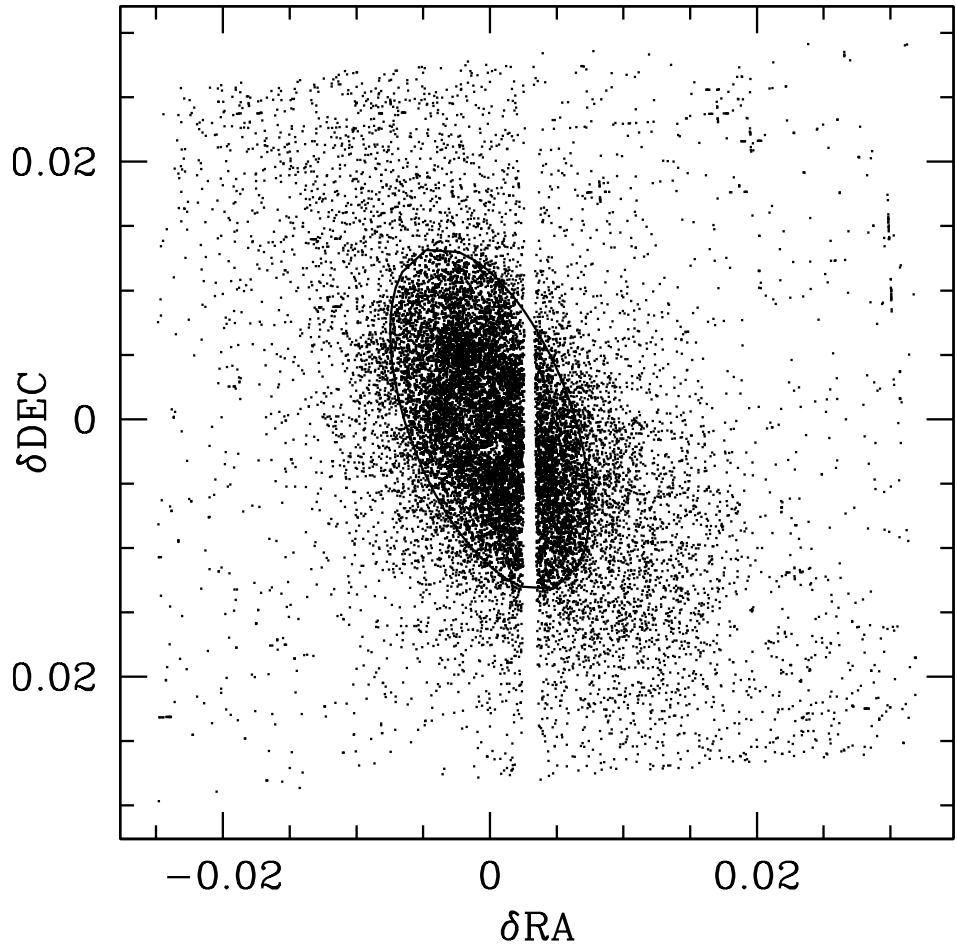,width=0.8\textwidth}}
\caption{\label{fig:fig6}Projected spatial distribution of detected  
objects around UGC 3755 from ACS observations.}
\end{figure}

\begin{figure}
\centerline{\psfig{figure=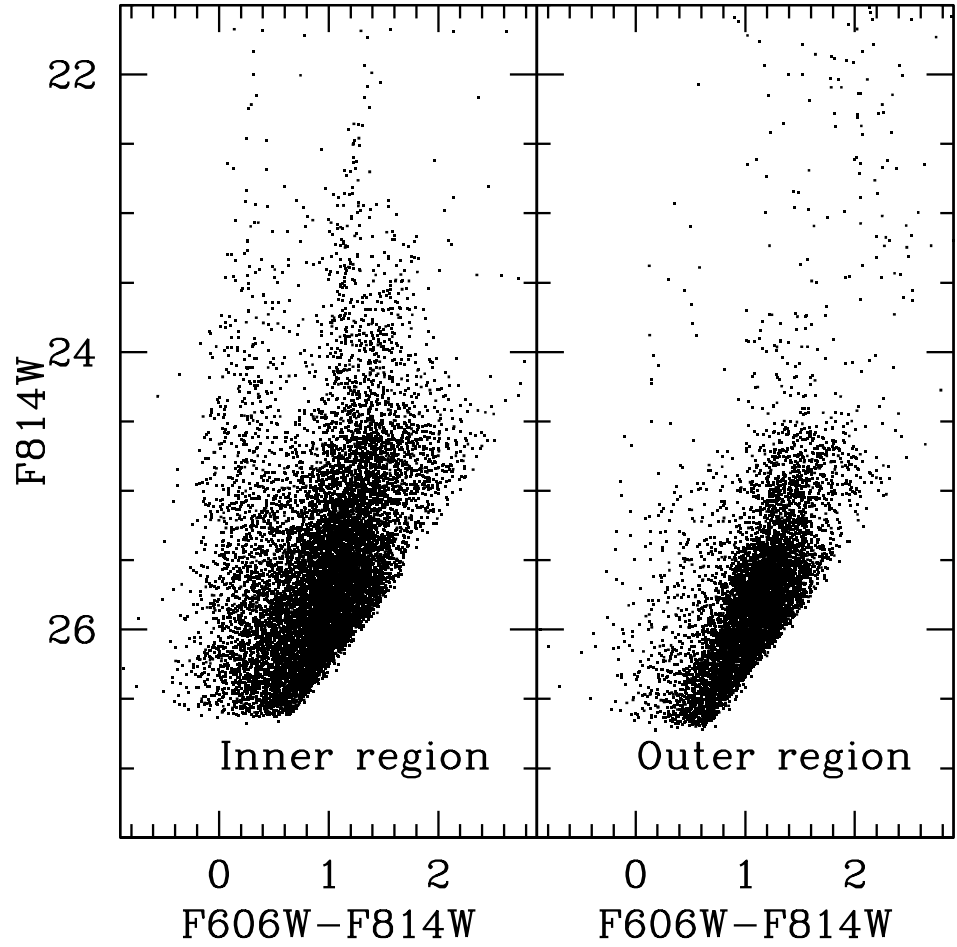,width=0.8\textwidth}}
\caption{\label{fig:fig7} Left panel: CMD of UGC3755 in the region  
inside the ellipse of Figure \ref{fig:fig6}. Right panel: CMD of  
UGC3755 in the region outside the ellipse.}
\end{figure}

When the TRGB is very close to the photometric limit, and the RGB is  
intrinsically contaminated by a significant population of  
intermediate-age or young stars, the detection of the tip can become  
very difficult or impossible. In these cases, knowledge of the  
spatial distribution of different stellar populations can be  
of great help in reducing the RGB contamination. As an example,  
Figure \ref{fig:fig6} shows the projected spatial distribution of all  
the detected objects around the galaxy UGC~3755, chosen from 
our ACS program 10210. The ellipse superimposed to the data has a major  
axis 0.014 degrees ($\sim$ 50 arcsec), an ellipticity 0.9, and a  
position angle 20 degrees. Left and right panels of Figure \ref  
{fig:fig7} show the corresponding CMDs for the regions inside and  
outside the ellipse, respectively. In this galaxy, as frequently occurs,
young  
and intermediate-age stellar populations tend to be more centrally  
concentrated.  As a consequence, the right panel of Figure \ref 
{fig:fig7} shows a much cleaner RGB, with very little contamination  
(note how the main sequence and supergiant sequences almost  
disappear). We could not obtain a satisfactory ML fit to the RGB tip 
with the CMD for
the whole galaxy, but our ML procedure successfully measures the tip 
on the spatially filtered
diagram, producing a value $I_{TRGB}=25.47\pm0.06$ ($D=7.4$~Mpc
with 0.17 reddening). 
This filtering procedure can  
successfully be applied to improve the precision of the TRGB
determination as long as the number of stars in the first magnitude   
below the tip does not fall below $\sim 100$.

\section{Summary}

Let us briefly summarize the results of our study. The main aim of this work is 
the improvement of the TRGB estimation accuracy and reliability closer to the 
photometric limit.
The necessity of a consistent methodology became evident during the analysis 
of our large stellar
photometry database of HST/WFPC2 and ACS observations of nearby  dwarf
galaxies. The resulting procedure is a logical enhancement of previous work.
The maximum likelihood algorithm is augmented with 
detailed modeling of 
photometric errors and the completeness function. The diverse testing of our 
method shows that it is 
reliable, has good accuracy, and does not introduce any systematic errors. 
This technique is
especially useful in cases where the TRGB approaches the photometric limit 
and/or the RGB is poorly populated in the CMD.  

The current study allowed the slope of the red giant branch luminosity
function to be a free parameter in the maximum likelihood fits.  However, 
in cases where the TRGB is within $\sim 1$~mag of the faint cutoff of the
photometry the determination of the slope $a$ is unstable.  Worse, there
is a coupling between the slope and the location assigned to the TRGB.
Because of the onset of incompletion, the ML fitting can spuriously chose
a very flat or even negative slope for the LF and simultaneously prefer
a value for the TRGB that is too faint.  These points were confirmed with
fitting to artificial CMD where the TRGB was precisely known.  If the 
TRGB is well removed from the faint limit then it is possible to solve
for the red giant LF slope with good accuracy.  Inevitable, the slope is
near to $a=0.3$ and the TRGB can be accurately measured assuming this
value for the slope.  

We further found that the slope $c$ describing the asymptotic giant branch
luminosity function should be constrained.  The slope of the AGB LF is less
predictable.  However, if left
free, nonphysically large positive values for the AGB slope can mask the 
full amplitude of the
TRGB jump.  Experience tells us that the AGB slope lies in the range 
$0.2<c<0.4$.

Presently, we make ML fits using two alternative constraints on LF slopes.
With the first alternative, the slopes for both the RGB and AGB are set at 
$a=c=0.30$.
With the
second alternative, the values of the slope parameters $a$ and $c$ are 
free but constrained by priors $a^{\prime} \pm \sigma_a =0.30 \pm 0.07$ and 
$c^{\prime} \pm \sigma_c =0.3 \pm 0.2$ where the primed values are the 
expectation slopes and $\sigma_a$ and $\sigma_c$ 
are the rms deviations around the expectation values found from 18 well 
observed LF.  The preferred approach is the second alternative if
there is a proper understanding of prior values and rms deviations and if
the incompletion function is being properly formulated.  We are considering
the two alternative formulations in order to test for systematics once
distance measures become available for the ensemble of galaxies 
suitably observed by HST.

The subject of the present study is just
the detection of the TRGB in a color--magnitude diagram and the 
assignment of errors to the detection. 
The important error
sources associated with calibration and extinction will be considered in
Paper~II. These potential
sources of systematic error can contribute up to $5-10\%$
uncertainty to the distance estimation \citep{bel}.
Paper~II will be concerned with the absolute value of the TRGB in $I$ and
HST flight F814W passbands, the effects of metallicity in the absolute
calibration, the effects of reddening, and the degree of agreement between
the TRGB distance determinations and other methodologies.

\acknowledgements
We acknowledge use of the Hubble Space Telescope data archive at STScI, which
is operated by the Association of Universities for Research in Astronomy, Inc.
under NASA contract NAS 5-26555 and support from grants associated with HST 
programs AR-9950 and GO-9771, 10210, 10235.
DM and LM are grateful to the Department of Astronomy, University of Maryland
and Institute for Astronomy, University of Hawaii for hospitality during this work.
This project also was supported by the Russian Foundation for Basic
Research grant 04--02--16115. DIM acknowledge support from INTAS grant
03--55--1754 and from the Russian Science Support Foundation. DM is also grateful to 
Observatoire de Lyon for hospitality during my visits.

{}

\newpage

\begin{table}
\caption{The TRGB detection results for NGC~2366}
\medskip
\begin{tabular}{lcccccc}
\hline
Field      &  Filters      & Exp. (sec) & $I_{TRGB}$               & 
$a$                     &  $b$                    &  $c$ \\ \hline
Inner (sum)& F555W, F814W  & 6700, 4100 & 23.55 $^{+0.04}_{-0.03}$ & 
0.40 $^{+0.01}_{-0.02}$ &  0.25 $^{+0.03}_{-0.03}$ & 0.54 $^{+0.07}_{-0.07}$ \\
Inner (a)  & F555W, F814W  & 1300, 1300 & 23.57 $^{+0.03}_{-0.04}$ &
0.38 $^{+0.02}_{-0.02}$ &  0.28 $^{+0.04}_{-0.03}$ & 0.48 $^{+0.07}_{-0.07}$ \\
Inner (b)  & F555W, F814W  & 1400, 1400 & 23.56 $^{+0.04}_{-0.04}$ &
0.38 $^{+0.02}_{-0.02}$ &  0.25 $^{+0.03}_{-0.03}$ & 0.55 $^{+0.07}_{-0.07}$ \\
Inner (c)  & F555W, F814W  & 1400, 1400 & 23.62 $^{+0.04}_{-0.03}$ &
0.35 $^{+0.02}_{-0.02}$ &  0.23 $^{+0.03}_{-0.03}$ & 0.65 $^{+0.07}_{-0.06}$ \\
Outer      & F606W, F814W  & 1400, 2000 & 23.56 $^{+0.06}_{-0.06}$ &
0.50 $^{+0.05}_{-0.04}$ &  0.39 $^{+0.09}_{-0.09}$ & 0.47 $^{+0.22}_{-0.20}$ \\ \hline
\end{tabular}
\end{table}

%
%

\end{document}